\documentclass{article}

\usepackage[ruled]{algorithm2e}
\usepackage{amsfonts,bm,enumerate}
\usepackage{amsthm,amsmath,amssymb,amsfonts}
\usepackage{mathtools,xfrac}
\usepackage{color,xcolor}
\usepackage{graphicx,epsfig,tikz}
\usepackage{caption,subcaption}
\usepackage{enumerate}
\usepackage{pdfsync,array}
\usepackage[normalem]{ulem}
\usepackage[title]{appendix}

\usepackage[sort]{natbib}

\usepackage[margin=1in]{geometry}
\usepackage{verbatim}
\usepackage{makeidx,latexsym}
\usepackage[colorlinks=true,citecolor=blue]{hyperref}
\usepackage{url}
\usepackage[capitalize]{cleveref}
\crefformat{equation}{(#2#1#3)}
\crefrangeformat{equation}{(#3#1#4)--(#5#2#6)}

\usepackage{authblk}
\usepackage[in]{fullpage}
\makeatletter
\def\imod#1{\allowbreak\mkern10mu({\operator@font mod}\,\,#1)}
\makeatother
\usepackage[compact]{titlesec}
\usepackage{booktabs}
\usepackage{setspace}

\theoremstyle{plain}
\newtheorem{theorem}{Theorem}[section]

\newtheorem{lemma}[theorem]{Lemma}
\newtheorem*{lemma*}{Lemma}
\newtheorem{proposition}[theorem]{Proposition}
\newtheorem*{proposition*}{Proposition}

\newtheorem*{observation*}{Observation}

\theoremstyle{definition}

\newtheorem{assumption}{Assumption}[section]

\theoremstyle{remark}

\newtheorem*{remark*}{Remark}

\newtheorem*{example*}{Example}

\allowdisplaybreaks

\definecolor{gold}{rgb}{0.85,0.65,0}

\def\beq{\begin{equation}}
\def\eeq{\end{equation}}

\def\fnote#1{\footnote}

\newcommand{\grad}{\ensuremath{\nabla}}

\def\cG{{\cal G}}

\def\cS{{\cal S}}
\def\cT{{\cal T}}

\def\cV{{\cal V}}

\newcommand{\bbE}{\mathbb{E}}

\newcommand{\bbN}{\mathbb{N}}

\newcommand{\bbP}{\mathbb{P}}

\newcommand{\bbR}{\mathbb{R}}

\newcommand{\bbZ}{\mathbb{Z}}

\DeclareMathOperator*{\argmax}{arg\,max}

\DeclareMathOperator{\Poisson}{Poisson}

\DeclareMathOperator{\NB}{NB}
\DeclareMathOperator{\GammaDist}{Gamma}

\DeclareMathOperator{\Conv}{Conv}

\def\log{\mathop{{\rm log}}}

\DeclarePairedDelimiter\abs{\lvert}{\rvert}%

\makeatletter
\let\oldabs\abs
\def\abs{\@ifstar{\oldabs}{\oldabs*}}
\makeatother

\title{Poisson Inventory Models with Many Items: An Empirical Bayes Approach}
\author[1]{Edward Anderson}
\author[2]{Nam Ho-Nguyen}
\author[2]{Peter Radchenko}
\affil[1]{Imperial College London, United Kingdom}
\affil[2]{The University of Sydney Business School, Australia}
\date{}

\begin{document}
\maketitle

\begin{abstract}
We consider inventory decisions with many items, each of which has Poisson demand. The rate of demand for individual items is estimated on the basis of observations of past demand. The problem is to determine the items to hold in stock and the amount of each one. Our setting provides a natural framework for the application of the empirical Bayes methodology. We show how to do this in practice and demonstrate the importance of making posterior estimates of different demand levels, rather than just estimating the Poisson rate.  We also address the question of when it is beneficial to separately analyse a group of items which are distinguished in some way. An example occurs when looking at inventory for a book retailer, who may find it advantageous to look separately at certain types of book (e.g. biographies). The empirical Bayes methodology is valuable when dealing with items having Poisson demand, and can be effective even with relatively small numbers of distinct items (e.g. 100). We discuss the best way to apply an empirical Bayes methodology in this context, and also show that doing this in the wrong way will reduce or eliminate the potential benefits.
\end{abstract}

\bigskip

\section{Introduction}

The fundamental problem for inventory management is to adjust inventory levels to achieve the right trade-off between the costs of holding inventory and the costs of a stockout. This requires knowledge of the distribution of demand, which is obtained through looking at the history of demand for this product. There are bound to be errors in our estimation of the demand distribution. In most cases demand will vary over time and so we will want to estimate on the basis of the most recent data, which will lead to additional uncertainty about the pattern of demand. There has been an interest in this area right from the beginnings of inventory theory in the 1950's and we briefly review some of this literature in the next section.

The problem we address shares with this previous work the characteristic that inventory decisions need to be made without full knowledge of the distribution of demand, but this is in the context of large numbers of different products or items. We are interested in how much inventory to hold for each of a large number of different items when each item has demand which occurs as a Poisson process, and where the rate parameter for each item is estimated on the basis of past demand.

Where demand occurs from a large number of independent customers, then the Poisson model is often appropriate. The conditions we need for this model are that there is a population of customers each of whom in each period makes an independent decision whether or not to purchase, and a customer will purchase at most one item. If the population is sufficiently large, with a correspondingly small probability of purchase by any individual, then we can ignore the complications around the modelling of repeat purchases by the same customer. Effectively, using a Poisson demand distribution is like using a queuing model for the arrival of customers each of whom purchases just one item.

We will assume that demand for different items is independent and so it is legitimate to ask how information about the demand for other items can be relevant. We will adopt an empirical Bayes approach, but before discussing this we should observe that there are good reasons to believe we can improve on the na\"{i}ve approach of taking the observed rate of demand as our best estimate of the underlying Poisson rate parameter.  Clearly there will be some items that by chance have higher than expected demand and others that have lower than expected demand. If we consider the item with the highest observed demand we can guess that it is likely to be in the former group and we may choose to reduce our estimate of the underlying rate parameter for this item. Similarly we can guess that an item where we observe zero demand is likely to have a Poisson parameter that is higher than zero.

These types of argument can be made more precise if we have a prior distribution for the rate parameter: for an individual item we can compare the observed demand with this prior and use Bayes to update our estimate of the rate parameter. The empirical Bayes approach does not require a prior distribution to be specified in advance, but instead makes use of the entire set of observations to make deductions about the prior. The empirical Bayes approach is particularly well suited to this application of independent Poisson data, and goes back to the work of Herbert Robbins published in 1956 \citep{lai1986contributions}. We will make use of Robbins formula and extensions in our development.

Our application focuses on the case with large numbers of different items each of which has a Poisson demand process. This type of problem occurs quite frequently in practice. Problems in which there are a very large numbers of items many with low levels of demand, occur in an online retailing environment where it is typical to have a long tail of items which are ordered infrequently (e.g \citet{baldauf2024cut}). Problems of the same type have also been studied in the context of spare parts inventory. The demand in such circumstances corresponds to failures in installed components and is usually modelled either as a Poisson process or compound Poisson.

The question of how data from many different items can be leveraged to make better decisions than by treating the items independently has been considered by many authors.  An important recent paper (which also considers inventory problems) is \citet{gupta2022data}.  They have shown how a shrunken-SAA approach, which has some similarities to empirical Bayes, can achieve significant improvements over a standard SAA approach.

In an inventory setting it is common to assume that for each item there is a set of observations from an unknown distribution of demand, with the optimal decision on a reorder amount depending on the true underlying distribution. This is the framework used  in \citet{gupta2022data}. However this approach will not work well when there is low demand and limited data. Rather than a sequence of demand observations we may have a single observation of the total demand over the last replenishment period - with this demand often being very small, or perhaps zero. Even when we have data on demand over a number of periods we may not want to go back more than a handful, because of possible changes in the underlying levels. Knowing that the demand is drawn from a Poisson distribution implies that the total demand level over the entire period is all the information that is needed. There is no clear way to make a direct comparison between our parametric model with a single observation for each item and the non-parametric approach of  \citet{gupta2022data} involving multiple demand observations.

In this paper we give a thorough treatment of the use of empirical Bayes in a newsvendor-like context where there are many items each of which has Poisson demand. Our main contributions are as follows:
\begin{itemize}
    \item We show through an asymptotic analysis that a naive approach in which we ignore the demand information from the other items can do very badly.  Moreover, even if we use an empirical Bayes approach we can still go very wrong by simply estimating the Poisson parameter for each item and and then determining the order quantity.
    \item We make numerical tests in this setting  of two competing approaches to empirical Bayes. Choosing the right method makes a difference.
    \item We use simulated data to check the effectiveness of our empirical Bayes methodology in different settings. We show that it can be very effective even with relatively small numbers of items (say 100). We also give an illustrative application on a set of real data from a publisher.
    \item We address the question of whether, given a large number of items, it will be better to split them up and analyse groups of similar items separately. We investigate how this decision should be made, and demonstrate on our simulated datasets that the benefits from doing this are likely to be small.
\end{itemize}

The paper is organized as follows. In the next section we review the literature, before setting out the inventory model we will explore in \cref{model}. \cref{Asymptotics} makes some comparisons between different approaches in the limiting case where the number of items is so large that probability estimates can be made precisely. The computational choices that need to be made when applying empirical Bayes in practice are discussed in \cref{sec:estimation}, and numerical results are given in \cref{sec:numerics}. Finally in \cref{sec:grouping} we discuss the question of whether the data should be grouped together according to characteristics of the items. The proofs of all our theoretical results are given in Appendix~\ref{app:asymptotics}.

\section{Literature review} \label{literature}

The problem of designing inventory systems in the absence of information on the distribution of demand has been addressed in various ways. Perhaps the simplest approach is to take the historical record of demands as the empirical distribution and then find the corresponding optimal solution. For example in a newsvendor problem in which we have observed 10 samples of demand we optimize for a demand distribution that takes each of these 10 values with equal probability. This is called a sample average approximation (SAA) approach and it will do well with large enough samples. Bounds on the performance of SAA in this context are discussed by \citet{levi2007provably,levi2015data,lin2022data}. 
It may be possible to improve on a straightforward SAA approach through consideration of the optimisation estimation components concurrently. \citet{liyanage2005practical} show how this is possible when demand has an exponential distribution. A different approach is taken by \citet{besbes2023big} who track the worst case performance of the SAA policy for newsvendor problems against changes in sample size.

An alternative is to use a Bayesian approach in which a prior belief about the distribution of demand is sequentially updated as new information becomes available.  \citet{azoury1985bayes} adopts this method to analyse a periodic review policy where demand comes from a distribution with known functional form but with unknown parameters. This work is based on some early work by  \citet{scarf1960some} and \citet{iglehart1964dynamic}. Other papers that explore the implications of this approach are \citet{lovejoy1990myopic,kamath2002bayesian}. 

A third approach is to use ideas of robustness to ensure that the inventory policy is not unduly influenced by the choice of functional form for the demand distribution. We may apply a min-max approach based on knowledge of the mean and standard deviation of the demand distribution, see \citet{gallego1993distribution} - this is an idea which also goes back to \cite{scarf1957min}. However, rather than use this moment-based approach, it is more appropriate when dealing with uncertain demand to use robust optimisation with an uncertainty set based on the data observed, using some distance measure in the space of distributions. For example, \citet{klabjan2013robust} use an uncertainty set derived from a chi-square goodness-of-fit test. \citet{bertsimas2006robust,solyali2016impact} consider a robust approach with budgets of uncertainty to find good inventory policies in an environment with stochastic demand in discrete time, and where there is a fixed ordering cost as well as inventory costs.  \citet{perakis2008regret} extend these ideas by using minimax regret rather than minimax cost.

There are many examples of empirical research on inventory problems with large numbers of items. For example, \citet{bradford1990bayesian} look at a case where a firm sells poster art. \citet{nenes2010inventory} describe a similar problem faced by a Greek commercial enterprise. A number of researchers have considered similar problems in the context of spare parts inventory. The classic problem of spare parts inventory involves a relatively large number of field locations or bases where parts are held near to the customer and these locations are resupplied from a central warehouse or depot (see e.g. \citet{hadley1961model,simon1971stationary}). 
In many of these models there is the possibility of moving items from one location to another. Without this, we could think of the different locations as constituting different items with the allocation between locations mirroring the idea of an allocation of limited warehouse space or inventory budget to different items.

In this paper we will apply the ideas of empirical Bayes to the inventory problem in this context. The literature on empirical Bayes is extensive and we will not attempt a complete review. Herbert Robbins introduced the term empirical Bayes and gives a discussion of some of these ideas in \citet{robbins1964empirical}. The key idea is to make use of a number of observations, $X_1, X_2, \ldots, X_n $, where the observation $X_i$ is drawn from a distribution $\ell_{\theta_i}$ parameterized by an unobserved $\theta$. The parameter $\theta_i$ is itself drawn from an unknown distribution $g$. However, even though $g$ is unknown, we may estimate it by considering the complete set of observations $X_i$. We then use this estimate of the prior $g$ in a Bayesian approach to make a better estimate of parameter $\theta$ given a new observation $X$. But this outline of how empirical Bayes works omits a key observation: we can typically apply the appropriate Bayes formula for $\theta$ without making an explicit estimate of $g$, but instead make use of the (marginal) distribution of $X$ estimated directly from the observations  $X_1, X_2, \ldots, X_n$.  This idea lies behind both Robbins formula \citep{lai1986contributions} when each $X_i$ is Poisson, and Tweedie's formula that applies more generally to exponential families \citep{efron2011tweedie}. 

It is fair to point out that there have been differences of view around empirical Bayes, see for example the discussion in \citet{carlin2000empirical}. There have been many studies in which an empirical Bayes approach has been found to be effective, but empirical Bayes is no panacea and can fail if applied without sufficient care. \citet{efron2014two} gives a useful discussion of different approaches to empirical Bayes and some of the difficulties that can arise, while \citet{brown2013poisson} does something similar for the Robbins formula approach for the Poisson. Both of these papers discuss the importance of some type of smoothing or regularisation to improve the performance of the method.

There have been surprisingly few papers that use empirical Bayes approaches within an inventory setting. There are two possible ways in which empirical Bayes could be applied. The first is to use the time series of observations for demand for a single item, as the basis for the method \citep[e.g.][]{samaniego1999improving}. Thus, we model demand as a random variable drawn from a parameterised distribution where the parameter varies randomly from one period to the next, but is drawn from an underlying distribution on the parameter set.  This approach has been used by \citet{ye2022empirical} to determine an inventory policy for spare parts that have extremely complex demand patterns. 

Another approach, which is the focus of our work, is to look at the demand pattern for a large number of items, and use empirical Bayes to leverage the commonality between different products. In the simplest empirical Bayes approach, an assumption is made on the form of the prior for the distribution of the parameter $\theta$, say, with the data being used to estimate the parameters for this prior - which in turn is then used for a Bayesian estimate of the unknown $\theta$ associated with the observation. This is the approach described in \citet{morris1983parametric}. An example of the use of this in an inventory context is \citet{aronis2004inventory} who estimate a posterior distribution for the rate parameter $\lambda$ in a spare parts inventory context. They assume a Gamma distribution for the prior and update on the basis of failures observed. In a different context \citet{dolgui2008extended} apply a beta-binomial model to slow moving items where the binomial parameter is drawn from a common beta distribution with parameters estimated using maximum likelihood. 

The use of empirical Bayes does not require any assumptions on the form of the distribution of the parameter, and we can adopt the Tweedie/Robbins approach which works directly with the marginal. This more general approach is taken in the paper by Bradford and Suegrue who assume heterogeneous Poisson demand and apply the Robbins formula to make estimates of Poisson rates that are used in a base stock inventory system (\citet{bradford1997estimating}).

We will consider the methods that can be used to decide how to group together different items prior to carrying out the demand estimation.  This is also an issue in dealing with demand estimation in a retail setting and has been studied by \citet{cohen2022data}. They adopt a somewhat similar approach to us in looking at a hypothesis testing framework to make these decisions. 

\bigskip

\section{Inventory modelling with Poisson demand} \label{model}

We suppose that there are $n$ items for which there may be demand. Since $n$ is large the firm may not stock all of these items in a fulfilment centre. When facing demand for an item that is not stocked the firm will need to obtain the item from a central warehouse, or use some other supplier for that particular item. We may assume that the same options are available if demand is greater than the inventory held. For each item $i=1,2,\ldots,n$ there is a fixed cost $b_i$ to keep that item in stock, and a holding cost $c_i$ per unit time. Each sale of item $i$ generates revenue $r_i$ if there is sufficient stock for that sale.  We will assume that where demand exceeds the amount in stock, or where the item is not held in stock, then there are additional costs in obtaining the item leading to a reduced revenue of $\bar{r}_i$ for each unit of demand. In this case the firm receives at least $\bar{r}_i$ for each unit of demand for item $i$ and in determining an optimal policy we may, without loss of generality, subtract this from $r_i$. Thus there is no loss of generality in taking $\bar{r}_i=0$. An alternative model would have different revenue amounts for these two cases (of demand exceeding stock, or an item not being held in stock).

The firm observes the demand for each item over a time interval $[0,T]$ and needs to decide what items to stock and how many of each. Again for simplicity we will assume that demand is observed even if it is greater than the inventory held - so we do not consider censored data. We assume that there is a regular replenishment schedule, and normalise the time interval between replenishment to unit length. 

Since we assume a Poisson distribution for the individual customer demand arrivals, it is sufficient to just count the total demand over the time interval $[0,T]$ rather than paying attention to the particular times that demand occurs, for example by looking at demand on individual days. The total demand is a sufficient statistic for a set of independent Poisson random variables with the same rate parameter. 

Consider a single item $i$, and write $\xi_i$ for the demand over the replenishment interval. Suppose this has a known probability distribution with $\bbP \left[ \xi_i =k \right] =p_i(k)$, $k=0,1,2,\ldots$. Then the profit made given an inventory choice of $x_i>0$ is 
\[
r_i \bbE \left[\min\{x_i,\xi_i \}\right]-b_i-c_ix_i = r_i \sum_{k=0}^{x_i} k p_i(k) + r_i x_i \left( 1 - \sum_{k=0}^{x_i} p_i(k) \right) -b_i-c_ix_i,
\]
and is zero if $x_i=0$.

Using the standard newsvendor argument, if $x_i$ is greater than zero, it is chosen as $x_i^*$, which is the smallest $x_i$ value with 
\[
r_i \left( 1 - \sum_{k=0}^{x_i} p_i(k) \right) < c_i,
\]
i.e. it is the first $k$ with $F_i(k)>1-(c_i/r_i)$ when $F_i$ is the distribution function for $\xi_i$. This is the $1-(c_i/r_i)$ quantile for the discrete distribution for $\xi_i$. However the choice $x_i = x_i^*$ is only made if the profit is positive, which relies on $b_i$ not being too high. Thus we choose $x_i=0$ if 
\begin{equation}\label{condition x_i=0}
	b_i > (r_i - c_i) x_i^* - r_i \sum_{k=0}^{x_i^*} (x_i^*- k) p_i(k).     
\end{equation}

The problem is straightforward if we know the probabilities $p_i(k)$, and we now consider different possible estimation approaches for these probabilities. We will assume that the demand for item $i$ follows a Poisson process with unknown rate parameter $\lambda_i$. We suppose that a decision is required at time $T$ and in the time interval $[0,T]$ we observe demand of $X_i$ for item  $i$, so that $X_i \sim \Poisson(\lambda_i T)$.

A \emph{na\"{i}ve approach} would be appropriate if we had no prior information and there is just one item. Then observing the demand $X_i$ for item  $i$, will give an estimate for $\widehat{\lambda}_i=X_i/T$.  We can use this to generate the required probabilities directly from the Poisson distribution, so 
\begin{equation}\label{p_i value}
	p_i(k)=\widehat{\lambda}_i^{k} \exp ( -\widehat{\lambda}_i)/k!
\end{equation}

If we had a prior distribution for the value of $\lambda$ then we could use a Bayes approach to find the (posterior) distribution of $\lambda_i$ based on the observed demand $X_i$. Rather than using a prior, the empirical Bayes approach uses the observed demand values for all the other items. These observed demands form the basis for an estimate of the overall demand distribution from which we can make deductions about the distribution for $\lambda$ within the population of all items. We can think of this in the same way that we think of a prior distribution, essentially giving a Bayesian approach to estimate the posterior distribution of $\lambda_i$ given the observation $X_i$. 

The simplest way to proceed, which we call \emph{Plug-in Empirical Bayes}, is to calculate the expected value of $\lambda_i$ from the posterior distribution. This then becomes our estimate $\widehat{\lambda}_i$ from which we can generate values for $p_i(k)$ from (\ref{p_i value}). This approach allows the use of Robbins formula which works directly with the demand distribution, and avoids the need to make estimates of the underlying $\lambda$ distribution.

A second option, which we call \emph{Full Empirical Bayes}, is to use the full posterior distribution of $\lambda _i$ in order to generate a distribution for the demand. Implicitly this involves an estimate $g_i(\lambda \mid X_i)$ of the density function for the posterior distribution of $\lambda$ given the observation $X_i$, and then the appropriate integral version of (\ref{p_i value})
\[
p_i(k)=\int \frac{w^{k} \exp (- w)}{k!} g_i(w) \, \mathrm{d} w
\]
gives the probabilities that are needed.

\bigskip

\section{Asymptotic performance of na\"{i}ve and plug in approaches} \label{Asymptotics}

\noindent Before starting on a detailed analysis of the different approaches to computing posteriors we want to show that the approach we take can make a big difference. The na\"{i}ve approach can fail badly, and we will show that a plug in approach may also have very poor performance. We do this through an asymptotic analysis in which we allow $n$, the number of items, to tend to infinity.

An empirical Bayes approach is an approximation to an exact Bayesian analysis, in which we know the actual distribution of the rate parameter $\lambda$, and hence are able to determine the exact values for the distribution of $\lambda$ conditional on our observation of a certain level of demand. The empirical Bayes approach provides a closer approximation to this as the number of different items increase. And for any empirical Bayes approach, an upper bound on performance is given by what happens when the underlying distribution of the rate parameter $\lambda$, is known. In our discussion in this section we do not consider empirical Bayes directly; instead, we make comparisons with what could be achieved when the distribution of $\lambda$ is known (see \cite{brown2013poisson}, \cite{james2022irrational}  for examples of the full asymptotic analysis of related empirical Bayes methods).

We consider a version of the problem where each item has identical revenue per unit, which we normalise to $1$, so $r_i=1$, $i=1,2,\ldots,n$. Moreover we suppose that each item has the same fixed positive cost to keep that item in stock, so that $b_i=b$, $i=1,2,\ldots,n$.

Here we take $T=1$ for simplicity. Thus for each item $i=1,...,n$, observations~$X_i$ are generated from $\text{Poisson}(\lambda_i)$. The rates~$\lambda_i$ are generated from a ``prior'' on $(0,\infty)$.  Let $f$ denote the underlying marginal distribution (probability mass function) of the observations~$X_i$, so that $f(k)$ is the probability that if a new item is selected at random from the entire set of items we observe a demand of $k$ in a  period of length $1$. We also define
\[
R(k)=\frac{f(k+1)}{f(k)}.
\]
As we will discuss below the quantity $R$ occurs in Robbins formula which states that
\[
\bbE[\lambda_i \mid X_i=k]=(k+1)R(k).
\]

For the asymptotic results below, we need to assume that $f(k)$ goes to zero quickly. Specifically we assume that for every $\varepsilon >0$, $f(k+1) \le \varepsilon f(k)$ for $k$ large enough. This can be concisely expressed as:
\begin{assumption}\label{ass:Rk-small}
	$R(k) \rightarrow 0$ as $k\rightarrow\infty$.
\end{assumption}

We write $\lesssim$ and $\gtrsim$ for inequalities that hold up to positive universal multiplicative factors, i.e. $a(n) \lesssim b(n)$ means that there is some constant $K$ where $a(n)<K b(n)$ for all $n$.

Assumption \ref{ass:Rk-small} holds when the possible values for $\lambda_i$ are bounded (\emph{Bounded prior}) which implies that the tails of $f$ are Poisson and $R(k)\lesssim 1/k$. It will also hold when the $\lambda_i$ are drawn from a \emph{Gaussian prior}. This is established by the following Proposition, since  $\sqrt{\log(k)/k} \rightarrow 0$.
\begin{proposition}
	\label{half.norm.prop}
	If $\lambda_i$ are drawn from the Half-Normal distribution, then $R(k)\lesssim \sqrt{\log(k)/k}$.
\end{proposition}

The optimal allocation maximizes the expected total profit given the observed values $X_1,...,X_n$. Ignoring the item fixed cost~$b$, the optimal allocation for item~$i$ is the $(1-c_i)$-quantile for the conditional distribution $X_i^{\text{new}} \mid X_i$, where $X_i^{\text{new}}$ is independently generated from $\text{Poisson}(\lambda)$ after $\lambda$ is generated from the ``posterior'' distribution $\lambda_i \mid X_i$, using the observed value~$X_i$.

For different policies we will evaluate the expected total profit with respect to the distribution of new demands given the observed data. So these new demands occur according to a posterior distribution for $\lambda_i$ that arises from the true prior for $\lambda_i$ after conditioning on the observed demand. We will denote the expected total profit for the optimal allocation as $\Pi^*_n$ and denote the corresponding expected total profit for the na\"{i}ve allocation as~$\widehat{\Pi}_n$.

Assume that the $c_i$ are randomly (and independently) generated from some probability distribution on $[C_L,1)$, where~$C_L$ is allowed to be any fixed constant in $(0,1)$. The distribution may include points with positive weight; so this includes the case where $c_i=C_L$ for all $i=1,2,\ldots,n$. Given this distribution of~$c_i$ values we now establish that the naive allocation may perform very poorly. We will show that for large enough fixed cost $b$ there will be an average \emph{loss} made on each item, with the average loss per item being bounded below by some constant. Hence as the number of items increase to infinity the total expected losses also increase to infinity.

\begin{theorem}
	\label{asymptotics.naive.thm}
	Under \cref{ass:Rk-small}, there exists a positive constant~$B$ such that $b\ge B$ implies  
	$\widehat{\Pi}_n\lesssim -n$ with probability tending to one as~$n$ tends to infinity.
\end{theorem}

This result establishes that a na\"{i}ve allocation will be disastrous in the case of high fixed costs. In this case there are enough items which should not be stocked, but are stocked under the na\"{i}ve allocation as a result of chance high demand, that we end up with large losses. We should note that under the optimal policy  $\Pi^*_n\ge0$ by definition. Moreover, if either the (prior) distribution of~$\lambda$ is unbounded or the distribution of~$c_i$ satisfies some mild conditions involving~$b$, then we can show that $\Pi^*_n\gtrsim n$ (see~\ref{app:opt.profit} for further details).

The result is given for the case with $T=1$, but the role of the interval $T$ in this result is minor. If $X_i\sim\text{Poisson}(\lambda_iT)$ and~$T$ is fixed, then Theorem~\ref{asymptotics.naive.thm} still holds. But we can ask what happens when we collect more data by letting the interval $T$ get longer. Suppose that $X_i\sim\text{Poisson}(\lambda_iT_n)$ and~$T_n\rightarrow\infty$. If $T_n=o\big(\log(n)\big)$, then Theorem~1 still holds, but~$B$ will generally depend on~$T_n$. However we can show that if the prior on~$\lambda$ is bounded and $T_n=o\big(\log(n)\big)$, then Theorem~\ref{asymptotics.naive.thm} still holds with a fixed~$B$.

We have shown that the na\"{i}ve approach may perform very poorly; what is more surprising is that the plugin approach can also do very badly. The intuition behind our next result is that if there is uncertainty about the value of $\lambda$ and we plug in the probabilities associated with the expected value, then we will consistently underestimate the probability of more extreme outcomes for demand, so that we will underestimate the probability of getting a demand of zero and also underestimate the probability of getting high demands. For certain values of the parameters this will result in very poor performance. In the next result we show that it can be disastrous for particular (high) values of the cost parameters $c_i$.

We will denote the expected total profit for the plugin allocation as $\Pi^{\text{plug}}_n$. We assume that the fixed costs $b_i$ are zero and the variable costs are the same for each unit, thus $c_i=c$ for all $i$. We also assume that the support of the prior for $\lambda$ is a finite set that contains at least two points.

\begin{theorem}
	\label{asymptotics.plugin.thm}
	There exist values for the positive constant~$c$ such that  $\Pi^{\text{\rm plug}}_n\lesssim -n$ and $\Pi^{*}_n\gtrsim n$ with probability tending to one as~$n$ tends to infinity.
\end{theorem}

We can see by comparing Theorems \ref{asymptotics.naive.thm} and \ref{asymptotics.plugin.thm} that the same kind of disastrous behaviour can occur for the plug-in policy as for the naive policy, but for the plug-in this happens for quite specific values of the costs, whereas the bad behaviour for the naive policy happens for very general distributions of costs provided $b$ is high enough.

Looking in detail at the proof given in Appendix~\ref{app:proofs} makes it clear that we can weaken the assumptions even further if we wished. It is not necessary for fixed costs to be zero, or for variable costs to be all the same. What is required is that costs are chosen high enough that we only hold one item in inventory for the items that are most likely to have high demand, and we hold no inventory for the other items.
\bigskip

\section{Estimation}\label{sec:estimation}

It is helpful to summarise our model framework here. We use the following Bayesian model for observed ($X$) and future ($\xi$) demand of a particular item:
\begin{equation}\label{eq:demand-model}
	\lambda \sim g(\cdot), \quad X \mid \lambda \sim \Poisson(\lambda T), \quad \xi \mid \lambda \sim \Poisson(\lambda),
\end{equation}
where we write $g$ for the underlying distribution of the parameter $\lambda$, taken as a distribution over $\bbR_+$. The future demand $\xi$ and the previous observation $X$ are independent samples from the appropriate Poisson distributions.  Given the item's rate parameter $\lambda$, ideally we would like to solve the inventory problem for future demand:
\[ \max_{x \in \bbZ_+} \left\{ r \bbE[\min\{x,\xi\} \mid \lambda] - cx - b \bm{1}_{(x \geq 1)} \right\}. \]
In practice, however, we are never given $\lambda$, but instead we observe past demand $X$, thus we instead aim to solve the proxy problem
\[ \max_{x \in \bbZ_+} \left\{ r \bbE[\min\{x,\xi\} \mid X] - cx - b \bm{1}_{(x \geq 1)} \right\}. \]

To solve this problem, we wish to estimate the posterior probabilities $\bbP[\xi = k \mid X=s]$, even though we do not have access to the underlying distribution for $\lambda$, which is $g(\cdot)$. The empirical Bayes approach can be applied in two different ways, both of which make use of the complete observation set $\{X_i\}_{i \in [n]}$. The first approach estimates the marginal distribution for $X$, denoted $f_T(\cdot)$, and uses this directly (without estimating $g$). The second approach uses the observations $X$ to estimate the underlying distribution $g$ from which the posterior distribution can be calculated.

\subsection{Computing posterior probabilities}

We can express the posterior probability that we require in terms of expectations taken over the value of $\lambda$.
\begin{align*}
	\bbP\left[ \xi=k  \mid X=s \right] &=  \frac{\bbE \left[ \bbP[\xi=k, X=s \mid \lambda] \right]}{\bbE \left[ \bbP[X=s \mid \lambda] \right]} = \frac{\bbE\left[ \frac{\lambda^k \exp(-\lambda)}{k!} \frac{(\lambda T)^s \exp(-\lambda T)}{s!} \right]}{\bbE\left[ \frac{(\lambda T)^s \exp(-\lambda T)}{s!} \right]}.
\end{align*}
Thus we obtain
\begin{equation} \label{eq:post_prob_g}
	\bbP\left[ \xi=k  \mid X=s \right] = \frac{\bbE\left[ \lambda^{k+s} \exp(-\lambda(T+1)) \right]}{k! \bbE\left[ \lambda^s \exp(-\lambda T) \right]}.
\end{equation}
If we use the complete set of observations to estimate the distribution $g$ we can evaluate the required expectations in this expression to find these probabilities, and hence to make the optimal choice of inventory to hold for an item where the observed demand is $X=s$. We call this the ``$g$-modelling approach'', following the terminology of \citet{efron2014two}.

But there is another approach which works directly with the marginal of $X$:
\[ f_T(s) = \bbP[X = s] = \bbE \left[\bbP[X=s \mid \lambda] \right] = \bbE\left[ (\lambda T)^s \exp(-\lambda T) / s! \right]. \]
Thus from (\ref{eq:post_prob_g}), we can write
\begin{align*}
	\bbP\left[ \xi=k  \mid X=s \right] &=\frac{ \bbE\left[ (\lambda T)^{k+s} \exp(-\lambda(T+1)) \right]}{T^k k! s! \bbE\left[ (\lambda T)^s \exp(-\lambda T)/s! \right]}\\
	&= \frac{\bbE\left[ (\lambda T)^{k+s} \exp(-\lambda T) \sum_{j=0}^\infty \frac{(-1)^j \lambda^j}{j!} \right]}{T^k k! s! \bbE\left[ (\lambda T)^s \exp(-\lambda T)/s! \right]}\\
	&= \sum_{j=0}^\infty \frac{ (-1)^j \bbE\left[ (\lambda T)^{j+k+s} \exp(-\lambda T) \right]}{T^{j+k} j! k! s! f_T(s)}\\
	&= \sum_{j=0}^\infty \frac{(-1)^j (j+k+s)!}{T^{j+k} j! k! s! f_T(s)} \bbE\left[ (\lambda T)^{j+k+s} \exp(-\lambda T) / (j+k+s)! \right].
\end{align*}
So we get
\begin{equation} \label{eq:post_prob_f}
	\bbP\left[ \xi=k  \mid X=s \right] = \sum_{j=0}^\infty \frac{(-1)^j (j+k+s)! f_T(j+k+s)}{T^{j+k} j! k! s! f_T(s)}.
\end{equation}
We describe this as the ``$f$-modelling approach'', which is attractive because it does not require any estimation of the distribution of $\lambda$, and instead makes use of the marginal distribution from which we have a sample.  Thus the quantities $f_T(s)$ can be estimated directly from the proportion of items for which $X_i=s$. Notice, however, that in the $f$-modelling formula, there is an infinite sum with large factorials in both numerator and denominator of each term. In practice, we may expect this formula to be numerically unstable.

We also consider an alternative simpler approach which is to use the following \emph{plugin estimator}: given (an estimate of) $f_T$, we compute the posterior expectation of $\lambda$ as follows:
\begin{align*}
	\bbE[\lambda \mid X] &= \frac{\bbE \left[ \lambda \bbP[X \mid \lambda] \right]}{f_T(X)}\\
	&= \frac{\bbE \left[ \lambda \cdot (\lambda T)^X \exp(-\lambda T)/X! \right]}{f_T(X)}\\
	&= \frac{(X+1) \bbE \left[ (\lambda T)^{X+1} \exp(-\lambda T)/(X+1)! \right]}{T f_T(X)}
\end{align*}
Thus we obtain Robbins formula for the expected value of $\lambda$:
\begin{equation} \label{eq:post_expectation_f}
	\bbE[\lambda \mid X] =\frac{(X+1) f_T(X+1)}{T f_T(X)}.
\end{equation}
We then simply assume $\xi \mid X \sim \Poisson(\bbE[\lambda \mid X])$, so the posterior probabilities are approximated as
\[ \bbP[\xi=k \mid X] \approx \frac{\bbE[\lambda \mid X]^k \exp(-\bbE[\lambda \mid X])}{k!}. \]
Note that this approximation is not guaranteed to converge to the true posterior probability as we collect more data.

In the next two subsections we will consider in more detail effective methods for the estimation of $f_T$ and $g$, used for the two different approaches. Since our aim is to compare these approaches we will make every attempt to apply the most effective estimation in each case.

\subsection{Estimating the marginal $f_T(\cdot)$}\label{sec:estimate-f}

Recall that we observe past demands $\{X_i\}_{i \in [n]}$. We let $y_s$ be the number of times $X_i = s$ in our dataset. The distribution $f_T$ is exactly the marginal distribution of the observations. Therefore the simplest estimation is given by:
\[ \hat{f}_{T,n}(s) = \frac{y_s}{n} \]
which is also the sample average approximation. However, this has finite support, i.e., $\hat{f}_{T,n}(s) = 0$ whenever $s > \max_{i \in [n]} X_i$, and we know that the marginal for $X$ is a mixture of Poisson distributions, so $f_T(s) > 0$ for all $s$. Thus this is inconsistent.


To improve the estimate,
we will impose some parametric form $f_{\beta,T}(s) \approx f_T(s)$, and
then estimate the parameters $\beta$.
We will use a maximum likelihood estimator for $\beta$. Let $\cS$ be the set of unique observations in $\{X_i\}_{i \in [n]}$. Then the $\beta$ estimator is obtained by solving
\[ \hat{\beta}_n = \argmax_{\beta} \sum_{s \in \cS} \frac{y_s}{n} \cdot \log(f_{\beta,T}(s)). \]
The form of the parametric model we choose is $f_{\beta,T}(s) = \exp(\beta_0 + w(s)^\top \beta)$, which is based on Lindsey's method. This has been used previously in the empirical Bayes literature \citep{efron2011tweedie}. Specifically, the basis functions $w(s)$ are from a natural cubic spline, and we augment the estimation problem with two bespoke conditions derived from the specific Poisson inventory model we consider, namely the probability and monotonicity constraints. The maximum likelihood estimator, including the bespoke conditions, can be formulated and solved as an exponential cone program, thus is amenable to off-the-shelf solvers such as Mosek. We provide further details in the Appendix~\ref{app:estimation}.

\subsection{Estimating the prior $g(\cdot)$}\label{sec:estimate-g}

We now consider the alternative approach which is to estimate the underlying distribution of $\lambda$ (i.e. the distribution $g$). From this we can calculate the conditional probabilities that we use in determining the optimal policy. We describe this as the ``$g$-modelling approach''. We do this via a non-parametric maximum likelihood estimation (MLE). Recalling that $y_s$ is the number of times $X_i = s$ in our dataset, and $\cS \subset \bbN$ is the set of all observed values. We can write the likelihood of observing a single value $X=s$ when $g$ is the distribution of $\lambda$ as $\int \frac{ (\lambda T)^s \exp(-\lambda T)}{s!}\, \mathrm{d} g(\lambda)$. Thus the log likelihood for the complete set of observations is
\[
\sum_{s \in \cS} y_s \log\left( \frac{\bbE _g \left[ (\lambda T)^s \exp(-\lambda T) \right]}{s!} \right)
\]
where the subscript $g$ indicates that the expectation is taken with respect to the distribution $g$. Thus the non-parametric MLE model is
\begin{equation}\label{eq:non-parametric}
	\max_g \sum_{s \in \cS} y_s \log\left( \bbE _g \left[ (\lambda T)^s \exp(-\lambda T) \right] \right) - \sum_{s \in \cS} y_s \log(s!).
\end{equation}

The non-parametric MLE problem \eqref{eq:non-parametric} is infinite-dimensional over distributions $g$, but it can be shown to have a solution with only a finite number of points appearing in $g$. For simplicity we will assume that $T=1$ in this section, but the results carry over to the case where $T \ne 1$. For a positive measure $\tilde{g}$ over $\bbR_+$, denote
\[
v_{\tilde{g}}(s) = \int_\lambda \lambda^s \exp(-\lambda) d\tilde{g}(\lambda),
\]
and let $v_{\tilde{g}}$ be the vector $\{v_{\tilde{g}}(s) : s \in \cS\}$. When $\tilde{g}$ is a probability measure then $v_{\tilde{g}}(s) = \bbE _{\tilde{g}} \left[ \lambda ^s \exp(-\lambda ) \right]$. Dividing the component $v_{\tilde{g}}(s)$ by $s!$ will give the probability of observing a demand of $s$ when $\tilde{g}$ is the distribution of $\lambda$. If we define \[ \cV := \left\{ v_{\tilde{g}} : \tilde{g} \text{ is a positive measure over $\bbR_+$ such that $\int_\lambda d\tilde{g}(\lambda) \leq 1$} \right\} \subset \bbR_+^{\cS}, \]
then the following optimization problem over $\cV$ is equivalent to the non-parametric MLE problem \cref{eq:non-parametric}:
\begin{equation}\label{eq:non-parametric-finitedim}
	\max_{v \in \cV} h(v), \quad \text{where } h(v) := \sum_{s \in \cS} \frac{y_s}{n} \log\left( v(s) \right).
\end{equation}

It is clear that $h$ is concave, and \citet{Simar1976} shows that $\cV$ is convex and compact, thus \cref{eq:non-parametric-finitedim} is a finite-dimensional convex optimization problem.
\citet{Simar1976} further shows that \cref{eq:non-parametric-finitedim} will have a unique optimal solution $v^*$, with $v^*$ corresponding to $g^*$ having $\int_\lambda dg^*(\lambda)=1$ (i.e. $g^*$ is a probability measure).
\citet{Simar1976} also shows that the support of $g^*$ is finite, possibly containing $0$, i.e., $g^* = \sum_{j \in [r]} p_j \delta_{\lambda_j}$, where $\delta_{\lambda_j}$ is the Dirac delta measure at $\lambda_j$. Letting $s^* = \max_{s \in \cS} s$, the number of support points $r$ can be shown to satisfy the bound $r \leq \min\left\{ |\cS|, \left\lceil (s^*+2)/2 \right\rceil \right\}$,
and in fact there is a unique $g^*$ corresponding to $v^*$.

\citet{Simar1976} proposes a procedure to solve \eqref{eq:non-parametric}, which turns out to be equivalent to the fully-corrective variant of the \emph{conditional gradient (CG) method} applied to \eqref{eq:non-parametric-finitedim}; see \citet{Jaggi2013} for a description of the general CG method. The main idea is to solve a restricted version of \cref{eq:non-parametric}, and its corresponding finite-dimensional equivalent \eqref{eq:non-parametric-finitedim}, at each iteration $t$, where $g$ is restricted to have a fixed finite support of size $t$. Then, checking the optimality condition for \cref{eq:non-parametric-finitedim}, the size of the support is expanded by one atom each iteration until a convergence criterion is met. \citet{JagabathulaEtAl2020,JiangGuntuboyina2021} also employ the CG method for non-parametric estimation of mixing distributions, but not in the Poisson likelihood context. While \citet{Simar1976} does not provide any guarantees for the procedure, we provide a proof of convergence to an $\epsilon$-optimal solution to \cref{eq:non-parametric-finitedim} within $O(1/\epsilon)$ iterations by adapting the analysis of \citet{JagabathulaEtAl2020}; see \cref{thm:CG-convergence} in the Appendix~\ref{app:estimation}.

The advantage of the CG method in this context is that the solution $v_t$ maintained at each iteration $t$ has a concise representation in terms of its generating distribution $g_t$, and furthermore, $g_t$ has finite support with at most $t$ atoms. Thus, at any iteration $t$, computing any necessary functionals of the current $g_t$ (e.g., expectations) is straightforward.

A full description of the CG method to solve \eqref{eq:non-parametric-finitedim}, as well as precise statements of the convergence guarantee, is provided in the Appendix~\ref{app:estimation}.
\section{Numerical study}\label{sec:numerics}

We perform a numerical study to compare different approaches:
\begin{description}
\smallskip
	\item [The na\"{i}ve method] where for each item $i$ we estimate $\hat{\lambda}_i=X_i/T$, i.e., we simply use the demand observation with item $i$. We then optimize the inventory assuming that demand for item $i$ is distributed $\xi_i \sim \Poisson(\hat{\lambda}_i)$.\smallskip
	\item [The $f$-modelling method] where we estimate the marginal distribution $\hat{f}(\cdot)$ via maximum likelihood, together with additional constraints based on the problem structure (see \cref{sec:estimate-f} for details). This gives rise to two variants: (i) the \emph{plugin method}, where we use the estimated marginals $\hat{f}$ to compute a posterior estimate $\hat{\lambda}_i \approx \bbE[\lambda_i \mid X_i]$ for each item $i$, then optimize the inventory assuming that demand for item $i$ is distributed according to $\xi_i \sim \Poisson(\hat{\lambda}_i)$; and (ii) the \emph{full posterior method}, where we use the estimated marginals $\hat{h}$ to compute an estimate for the \emph{full} posterior distribution $\hat{p}_i(k) \approx \bbP[\xi_i=k \mid X_i]$. We then optimize inventory assuming that demand for item $i$ is distributed according to $\xi_i \sim \hat{p}_i$.\smallskip
	\item [The $g$-modelling approach] where we estimate the prior distribution $\hat{g}(\cdot)$ using a non-parametric MLE approach with the conditional gradient algorithm (see \cref{sec:estimate-g} for details). We then use $\hat{g}$ to compute posteriors $\hat{p}_i(k) \approx \bbP[\xi_i=k \mid X_i]$, and optimize inventory assuming that demand for item $i$ is distributed according to $\xi_i \sim \hat{p}_i$.
\end{description}

\subsection{Data generation}

We will conduct experiments on simulated data. Each instance of a problem is determined by the prior distribution for $\lambda$, as well as the revenue, fixed cost and holding cost values for each item.

We take the prior distribution $g(\cdot)$ for $\lambda$ as a Weibull distribution, with shape $\alpha$ and scale $\beta$ parameters, i.e., the distribution function is
\[ g(\lambda; \alpha, \beta) = \frac{\alpha}{\beta} \left( \frac{\lambda}{\beta} \right)^{\alpha-1} \exp\left( - (\lambda/\beta)^\alpha \right). \]
These are unimodal distributions with mode
\[ \max\left\{0, 1 - 1/\alpha\right\}^{1/\alpha} \beta, \]
so that higher values of $\lambda$ become more likely as $\beta$ increases. In our experiments, for each instance we fix $\alpha=1.8$ and set $\beta \in \{2, 3, 4, 5\}$. The mode for fixed $\alpha=1.8$ is $ \approx 0.6373 \beta$.

There is no particular significance to our reporting of results for the Weibull distribution. We have carried out other experiments with different distributions and the results are similar. Our choice of the Weibull distribution for the numerical experiments is made because it has the shape that we might expect in practice. With this choice of $\alpha$ there is a significant density that is near zero, implying a large number of with low demand per period. At the same time, there is a right hand tail giving a small number of items with large demand.

The revenues are set to be $r_i=1$ for each item $i$.  The costs $c_i$ are randomly generated $c_i \sim U(0.5, 0.9)$ for each item $i$. The fixed costs are set to be $b_i = 0.2$ for each item $i$.

Once we have generated the $\lambda_i$, we generate observations $X_i \sim \Poisson(\lambda_i)$ for each item. In summary, each instance is a set of tuples $\{ (\lambda_i, r_i, c_i, b_i, X_i) \}_{i \in [n]}$. We construct instances with various number of items $n \in \{50, 100, 250, 500, 1000, 5000, 10000, 25000, 50000\}$ and examine the effect as $n$ increases.

For each $\beta$, we generate 50 instances with $n=50000$ items according to the method above. Then, for each instance, nested subsets of these items are chosen of size $n=50,\ldots,25000$. We construct instances in this way to accurately capture the effect of increasing size.

We will also record for our problem instances the proportion of items held in inventory and the average number held conditional on this being more than zero. Both the proportion and the average number held are determined by a combination of $b$ and the scale parameter $\beta$. Increasing the fixed cost $b$ will lead to more items for which it is not economical to hold them in inventory, and increasing the average demand level through $\beta$ will increase the average number held. But these two effects are not independent since as $b$ increases more and more items with low demand will have inventory set to zero, and the average number held conditional on there being more than zero will therefore increase.

\subsection{Evaluation}

Each estimation method will construct an estimate of the marginal $\hat{f}$ or the prior $\hat{g}$ by \emph{only} using the observations $\{X_i\}_{i \in [n]}$ without knowledge of $\{\lambda_i\}_{i \in [n]}$. Then, based on $(r_i,c_i,b_i)$ and the constructed estimate, we compute an estimated inventory level $\hat{x}_i$ for each $i$. To evaluate the quality of our solution, we compare its performance against the best possible that can be achieved without knowledge of the true $\lambda$ value. To calculate this we compute the true posterior distribution for $\lambda_i$ given the observation $X_i$ using the true prior distribution $g(\cdot)$ for $\lambda$. We write $g^P_i$ for this distribution and $f^P_i$ for the corresponding marginal giving the probability of observing demand at different levels. This gives rise to the best possible inventory level $x_i^*$ which maximizes the profit obtained when $\lambda \sim g^P_i$, i.e. it maximizes
\[ \Pi(x;X_i,r_i,c_i,b_i)=\bbE_{X \sim f^P_i} \left[ r_i \min\{X, x\} \right] - c_i x - b_i \bm{1}(x > 0), \]
For each $n=50,\ldots,50000$, there are different 50 instances with $n$ items. For each instance, we record the average profit across the items. Then, we record the relative percentage gaps against the optimal average profit obtained using the true posterior distribution: $\frac{\text{average optimal profit} - \text{average profit of method}}{\text{average optimal profit}} \times 100\%$. Thus, smaller gaps indicate better performance.

\subsection{Results}

\cref{fig:profit-comparison} plots results for different $\beta$, comparing $n$ versus the average optimality gap.
We consistently see that $g$-modelling outperforms the plugin method, which in turn outperforms the na\"{i}ve method in \cref{fig:profit-comparison} for all $\beta$. We note that for $\beta=2$, some points have a relative gap of $>100\%$ thus are not shown on the plot. This is a result of the average profit being $<0$. In fact, for $\beta=2$, the na\"{i}ve method results in gap $>100\%$ for all $n$, indicating that its solutions are too aggressive, thus incurring the fixed cost even when demand is insufficient to recoup this. Further, we notice that the performance of the na\"{i}ve method is relatively insensitive to changes in $n$, whereas the other two methods improve as $n$ increases. This is expected, since more $n$ implies increasingly accurate estimations. For $\beta = 2,3,4$, the performance of the full posterior method seems to be between the plugin method and the na\"{i}ve method, while for $\beta = 5$ it is worse than the na\"{i}ve method.

\newcommand{\scalefactor}{0.5}
\begin{figure}[htb!]
	\begin{minipage}{.5\textwidth}
		\centering
		\includegraphics[scale=\scalefactor]{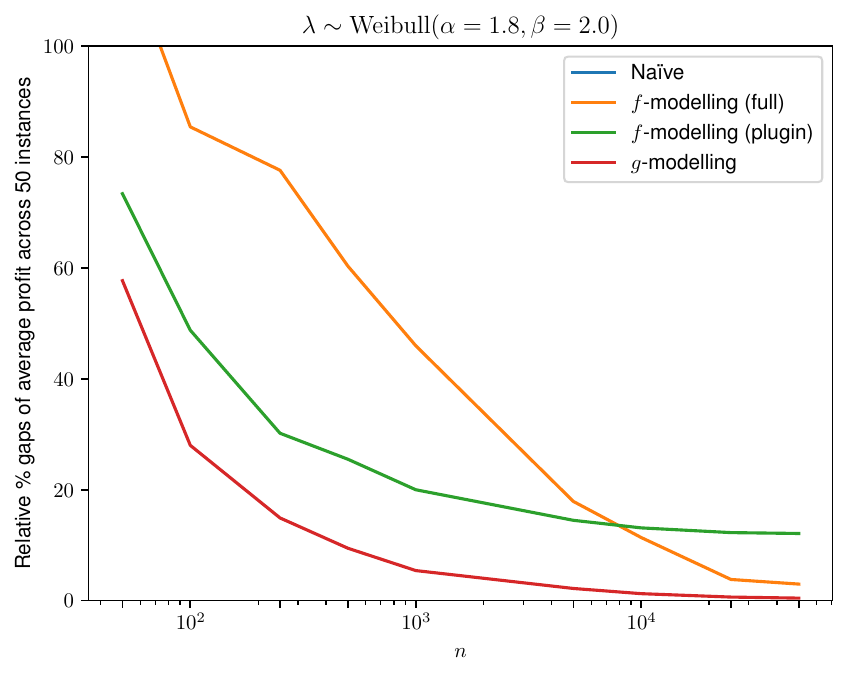}
		
\end{minipage}
	\begin{minipage}{.5\textwidth}
		\centering
		\includegraphics[scale=\scalefactor]{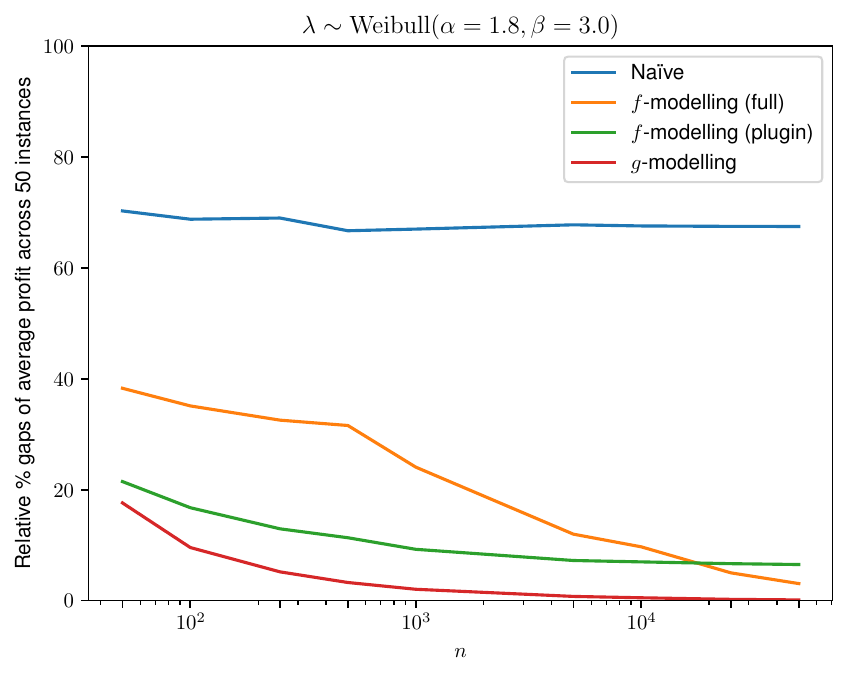}
	\end{minipage}
    \begin{minipage}{.5\textwidth}
	\centering	
        \small{$\beta=2$}
    \end{minipage} 	
    \begin{minipage}{.5\textwidth}
        \centering	
        \small{$\beta=3$}
    \end{minipage}
	\begin{minipage}{.5\textwidth}
		\centering \includegraphics[scale=\scalefactor]{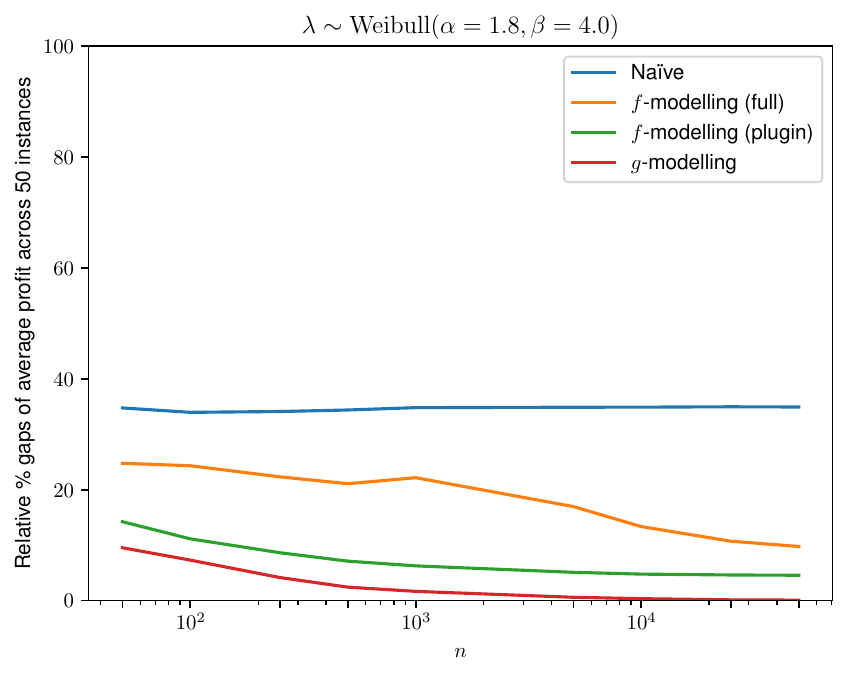}
	\end{minipage}
	\begin{minipage}{.5\textwidth}
		\centering \includegraphics[scale=\scalefactor]{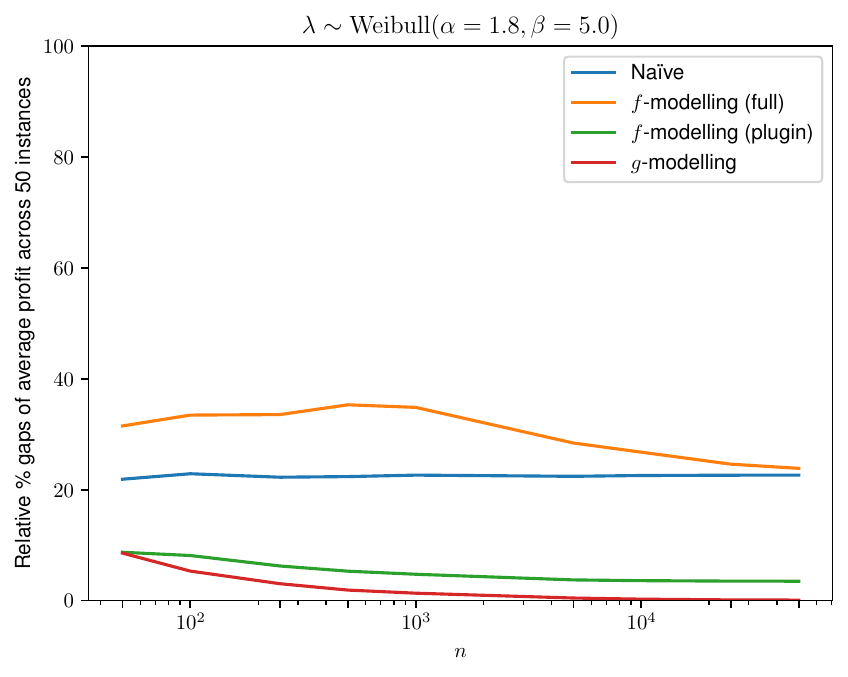}
	\end{minipage}
  \begin{minipage}{.5\textwidth}
	\centering	
        \small{$\beta=4$}
    \end{minipage} 	
    \begin{minipage}{.5\textwidth}
        \centering	
        \small{$\beta=5$}
    \end{minipage}

	\caption{Average percentage optimality gap versus number of items $n$ over 50 instances for $\alpha=1.8$ and different $\beta$.}
	\label{fig:profit-comparison}
\end{figure}

Conceptually, we would expect that the full posterior method would outperform the plugin method, yet this is not what is observed. We attribute the observed behaviour to numerical instability. Specifically, the full posterior method requires us to compute large factorials, which inevitably incurs numerical overflow, even when standard techniques are applied to facilitate division of large numbers. We explore this further in the Appendix~\ref{app:conjugate-prior}, where we tested \eqref{eq:post_prob_f} when given access to the exact marginal distribution under a gamma prior (which is conjugate to the Poisson likelihood, resulting in negative binomial marginal and posterior distributions).


Thus we see that full empirical Bayes, which requires estimates of the posterior probabilities, is computationally very demanding if an $f$-modelling approach is used. Indeed we do not have an effective approach to do this using $f$-modelling. Our earlier asymptotic result already suggested that full empirical Bayes would be worthwhile, these results confirm this and show that it is the $g$-modelling version of this approach that should be used.

\subsection{When does empirical Bayes perform well?}

We have shown that the $g$-modelling approach is effective across a range of sample sizes and for different scaling parameters in the Weibull. We will simply refer to this approach as empirical Bayes (EB) in our discussion. We now consider a wider range of parameter settings to investigate when we may expect to see a significant gain from the use of empirical Bayes for this inventory problem.

In these experiments we have continued to use the Weibull distribution with $\alpha=1.8$. The shape of this distribution with support on $(0,\infty)$ but with a small mode is suitable for this application, and we do not believe that changes in the distributional form will have a significant impact on our results. We have also continued to use costs $c_i$ that vary for each item (random between 0.5 and 0.9) with a constant revenue $r_i=1$. However we now consider variations in the fixed cost $b_i$ with this taking values in $\{0,0.1,0.2,0.3,0.4,0.5,0.6\}$ (but being the same for each item). We consider instances with $n \in \{ 100, 500, 1000, 2000, 10000, 15000, 20000 \}$ and Weibull scale parameter  $\beta \in \{2,3,4,5\}$.

Thus we have three main variations to consider, changes in the sample size $n$, changes in the underlying level of demand governed by the scale parameter $\beta$, and changes to the fixed cost $b$.

Our results when varying $n$ mirror those of \cref{fig:profit-comparison}. Changes in $b$ do not change the essential characteristics of these results and we see that the improvement in comparison with the na\"{i}ve approach rapidly increases as $n$ increases with most of the benefit available when $n=1000$ and no real additional benefit available for $n$ greater than 5000.

\begin{table} [hbt!]
	\centering
	\caption{Empirical Bayes results with different scale parameters ($n=100$)} \label{tab:gm-vs-naive-n=100}
	\begin{tabular}{ccccc}
		\toprule
		$\beta$ & Av. profit na\"{i}ve & Av. profit EB & Absolute improvement & Relative improvement\\
		\midrule
		2 & -0.04587 & 0.02432 & 0.07019 & \\
		3 & 0.03077 & 0.10171 & 0.07093 & 2.305 \\
		4 & 0.12913 & 0.19678 & 0.06765 & 0.524 \\
		5 & 0.26964 & 0.34045 & 0.07081 & 0.263 \\
		\bottomrule
	\end{tabular}
\end{table}

It is clear that both increasing the fixed cost $b$, and reducing the average demand by reducing $\beta$ will result in lower average profits. For either high values of $b$, or low values of $\beta$ the na\"{i}ve approach will often make losses. When we consider relative improvement over a naïve approach the benefits from using empirical Bayes are enormous for problems in which profitability is an issue. For example with $n=1000$, taking the average over all runs and $b$ values with scale parameter 2 (so low demand) gives an average loss of 0.04045, whereas using $g$-modelling reverses this and we get an average profit of 0.03016. Even with small values of $n$ something similar occurs. Table~\ref{tab:gm-vs-naive-n=100} shows the improvements from using empirical Bayes when $n=100$ and with different scale parameters. There are very large relative improvements and even when $\beta$ is 5 there is 26\% improvement from the use of empirical Bayes.

\begin{table} [hbt!]
	\centering
	\caption{Empirical Bayes results with different values of $b$ ($n=100$)} \label{tab:gm-vs-naive-b-variation}
	\begin{tabular}{ccccc}
		\toprule
		$b$ & Av. profit na\"{i}ve & Av. profit EB & Absolute improvement & Relative improvement\\
		\midrule
		0 & 0.24002	& 0.31186 &	0.07184 & 0.299 \\
		0.1 & 0.16901 & 0.23948 & 0.07047 & 0.417  \\
		0.2 & 0.11656 & 0.18567 & 0.06911 & 0.593 \\
		0.3 & 0.07701 & 0.14634 & 0.06933 & 0.900 \\
		0.4 & 0.04503 & 0.11503 & 0.07001 & 1.555 \\
		0.5 & 0.02117 & 0.09078 & 0.06961 & 3.288 \\
		0.6 & 0.00262 & 0.07153	& 0.06890 & 26.286 \\
		\bottomrule
	\end{tabular}
\end{table}

We get similar results when we consider variations in the fixed cost $b$. Table~\ref{tab:gm-vs-naive-b-variation} shows what happens for $n=100$, again we choose a low value of $n$ to show the extent of the improvement even when empirical Bayes does not have a very large sample to work with. For these results we average over all runs and values of $\beta$. The improvements are relatively large and even with $b=0$ amount to almost 30\%.

\begin{table} [hbt!]
	\centering
	\caption{Comparison of \% items held in stock and average numbers ($n=10000$, $\beta=3$)} \label{tab:gm-vs-naive-number-in-stock}
	\begin{tabular}{cccccccc}
		\toprule
		& \multicolumn{3}{c}{\% of items held in stock} & & \multicolumn{3}{c}{Av. number held in stock}\\
		$b$ & optimal & EB & na\"{i}ve & & optimal & EB & na\"{i}ve\\
		\midrule
		0 & 80.95 & 80.84 & 68.90 & & 1.750 & 1.763 & 2.621 \\
		0.1 & 60.75 & 60.84 & 57.13 & & 1.998 & 2.012 & 2.952 \\
		0.2 & 41.98 & 41.89 & 46.14 & & 2.368 & 2.391 & 3.357 \\
		0.3 & 29.82 & 30.05 & 37.76 & & 2.667 & 2.678 & 3.721\\
		0.4 & 22.17 & 22.24 & 31.07 & & 2.867 & 2.889 & 4.031 \\
		0.5 & 15.12 & 15.38 & 25.63 & & 3.186 & 3.198 & 4.349 \\
		0.6 & 10.68 & 10.89 & 21.60 & & 1.750 & 1.763 & 2.621 \\
		\bottomrule
	\end{tabular}
\end{table}

Next we look in more detail at where the improvement from empirical Bayes comes from. Table~\ref{tab:gm-vs-naive-number-in-stock} shows both the percentage of items held in stock and the average number held in stock conditional on this being greater than zero. We do this for the optimal policy and for the na\"{i}ve and EB approximations. These results are for $n=10000$, $\beta=3$, but the results are similar for other values of $n$ and $\beta$. We see that with the na\"{i}ve method there are substantially too high stock levels on average and, in cases where there is a high fixed cost $b$, too great a percentage of items are held in stock. Since these results are for a high $n$ value, the empirical Bayes approach achieves more or less the same numbers as the optimal policy.

\begin{table} [hbt!]
	\centering
	\caption{Profitability comparison ($n=1000$, $\beta=5$)} \label{tab:gm-vs-naive-profitbility}
	\begin{tabular}{ccccccc}
		\toprule
		& \multicolumn{3}{c}{Items held in stock} & & \multicolumn{2}{c}{Items not held in stock}\\
		$b$ & optimal & EB & na\"{i}ve & & EB & na\"{i}ve\\
		\midrule
		0 & 0.60040 & 0.59393 & 0.49929 & & -0.00676 & 0.00000 \\
		0.1 & 0.57549 & 0.56879 & 0.46316 & & -0.00390 & -0.00158 \\
		0.2 & 0.58196 & 0.57500 & 0.45391 & & -0.00379 & -0.00586 \\
		0.3 & 0.57421 & 0.56652 & 0.43371 & & -0.00263 & -0.01283\\
		0.4 & 0.57242 & 0.56437 & 0.41413 & & -0.00252 & -0.01837 \\
		0.5 & 0.56243 & 0.55369 & 0.38995 & & -0.00212 & -0.02347\\
		0.6 & 0.55933 & 0.54968 & 0.37169 & & -0.00173 & -0.02819 \\
		\bottomrule
	\end{tabular}
\end{table}

Finally we look separately at the profitability for those items held in stock in the optimal policy and the profitability for the other items. Table~\ref{tab:gm-vs-naive-profitbility} gives the average profit level under different approaches when $n=1000$ and $\beta=5$. For those items not held in stock in the optimal policy the profit is zero for the optimal policy, and hence this column is omitted. With higher $b$ values there are quite a lot of items that should not be held in inventory (64.8\% when $b=0.6$) and thus the losses made on those items could be significant. However, we see that average losses for these items are small for empirical Bayes. The losses are larger for the na\"{i}ve method, but even so are much smaller than the loss arising from not making the right choice of inventory amount amongst items that should be stocked.

\subsection{An example using real data}

To illustrate the methods in this paper, we explore performance on real data obtained from a small scale academic publisher. The publisher makes use of print on demand approaches to deal with titles with small numbers of sales over long periods of time. Sales are made through a web site. If there is no stock on hand of a book that is ordered it can be provided as a single copy printed on demand.  But the costs are higher and there is a delay before the book can be delivered. It is also possible for some titles, to print copies in advance of demand to ensure that stock is available, as happens already when the title is launched.  Printing in advance is cheaper per copy, but there are some fixed costs of delivery etc.  We will consider the decision needed for books that have exhausted their initial print run (so there is no stock available): is it worth printing in advance of future demand and if so how many should be printed?

%

We simplify the problem by looking at two years of sales: namely 2022 for model estimation and 2023 for validation. We performed some minor preprocessing, including reconciling duplicate sales and titles, and removing titles with discounted sale prices. After preprocessing, our data set consists of 178 titles with an average list price of \$50. The average number of sales for these titles is 1.8 and about half of the titles did not sell any copies during 2022.

We consider data from 2022 to determine the optimal decision. In reality decisions are needed at different times of the year as different titles run out of stock, but we suppose this review happens for every title at the start of 2023. Then we can look at the impact of our policy on overall profit for 2023, given the actual demand that occurs in 2023. This approach allows a fair comparison between the method we recommend and other approaches.

To be more precise, let $d_i$ be the number of sales for 2022 for book title $i$, and $s_i$ be the number of sales for book $i$ in 2023. Let $C_i$ be the list price of book $i$. We will use the following cost/revenue parameters: $c_{i}=0.08C_{i}$, $r_{i}=0.2C_{i}$, and $b_i = b$ is the same for every book, where we try $b=3,4,5,6,7$. The choices of values for these costs are quite uncertain. Our analysis is intended as illustrative, rather than a full case study.

We use $d_1,\ldots,d_n$ to estimate the distribution of future demand conditional on observing $d_i$ for each title $i \in [n]$ (where $n=178$). We then choose $x_i$ to optimize the (estimated) expected profit $r_i \bbE[\min\{\xi_i,x_i\} \mid d_i] - c x_i - b_i \bm{1}(x_i > 0)$, where $\xi_i$ is the future sales for book $i$.
We validate our decisions $x_i$ by computing the realised 2023 profit $r_i \min\{s_i,x_i\} - c_i x_i - b_i \bm{1}(x_i > 0)$. We report the total actual realised profit $\sum_{i =1}^n \left( r_i \min\{s_i,x_i\} - c_i x_i - b_i \bm{1}(x_i > 0) \right)$ for each method.

\cref{tab:book-data-results} provides some results for different methods below. Based on previous results in our simulation study, we elected to compare only $g$-modelling, plugin $f$-modelling and MLE approaches.
It is clear that the MLE method is the worst-performing one across all $b$, and the difference increases as $b$ increases. This reflects the result stated in \cref{asymptotics.naive.thm}. Furthermore, the full $g$-modelling method consistently outperforms the plug-in $f$-modelling method across all $b$. These results confirm that the empirical Bayes approach is very effective on real data, even with less than 200 items.

\begin{table}[ht]
	\centering
	\caption{Comparison of $g$-modelling versus MLE methods on book sales data. The two columns show the total profit for the estimated inventory solution: $\sum_{i =1}^n \left( r_i \min\{x_i,s_i\} - c x_i - b_i \bm{1}(x_i > 0) \right)$, where $s_i$ is the sales from 2023, while $x_i$ is the maximizer of the estimated conditional expectation on 2022 sales data.}\label{tab:book-data-results}
\begin{tabular}{lrrr}
\toprule
 & $g$-modelling  & Plugin  & MLE profit \\
 &  profit  &  $f$-modelling profit   &     \\
\midrule
$b=3$ & 418.518 & 353.172 & 279.540 \\
$b=4$ & 365.744 & 321.356 & 247.724 \\
$b=5$ & 359.356 & 272.156 & 198.524 \\
$b=6$ & 350.956 & 245.638 & 172.006 \\
$b=7$ & 347.560 & 243.838 & 170.206 \\
\bottomrule
\end{tabular}
\end{table}

\section{Grouping of items} \label{sec:grouping}

Using ean empirical Bayes approach typically involves pooling data across many items, under the assumption that each item’s underlying parameter is drawn from a common prior distribution. When we have no information characterising the different items then this makes sense, but in practice we frequently have some additional structure in the data that can be helpful. We will consider the appropriate trade-off when making use of covariates that allow us to separate out a subset of the total set of units. For example, in predicting book sales, when should we look separately at non-fiction books, or even break this down even further to historical biographies? Doing so will reduce the pool of data from which we are estimating the marginal or underlying distribution, but we will gain from the more consistent pattern of demand for a group of very similar titles. The general problem of incorporating covariate information into empirical Bayes is discussed by \cite{ignatiadis2019covariate} and \cite{efron2016empirical}.

Consider two sets of items (e.g., different book categories). It is only worthwhile to make inventory decisions using empirical Bayes separately for the two sets if there is a difference in the underlying distribution of demand (i.e. a difference in the $g$-distribution). The question can be posed in a different way: When some items belong to a restricted subset and we ignore this in determining an inventory policy, do we gain enough from the increased sample size to compensate for the difference in distribution between the restricted subset and the complete set of items? This trade-off between sample size and difference in behaviour also occurs when using a statistical test to determine whether two types of item have the same characteristics for their demand: either large differences or large sample sizes will be needed to be confident of a difference. This makes it natural in the empirical Bayes framework to make the decision to split or to combine two types of item based on the evidence that a subset of units (e.g., non-fiction titles) has a significantly different distribution of demand than all the other units. Hence we consider a hypothesis test with the null hypothesis that there is no difference between the distribution of demand observed in the subset and the distribution in the complete set of observations. Note that we adopt the hypothesis testing framework pragmatically, not rigidly. Thus, we might not be very confident that the restricted subset has different behaviour than the complete set of items, but this could still be enough to tip the balance in favour of treating the subset separately. So we do not expect that the decision to split will occur at some standard significance level of, say, 5\%.  In our inventory examples, we will explore the question of the significance level we might use in deciding to look at the restricted subset on its own.

There are several options for carrying out a hypothesis test in these circumstances. The simplest approach is to use a chi-square test of homogeneity on the basis of the marginals (the numbers of items with different levels of observed demand). However, there is an alternative approach that makes more direct use of $g$-modelling estimation of the parameter distribution, providing a natural link to a likelihood ratio test for differences between the nominated subset and all the other units. We have two nested models: the smaller one requires that the two subsets share the same set of $g_j$ parameters, and the larger one allows the two sets of parameters to be different. The test statistic is twice the logarithm of the ratio of maximized likelihoods (over the two models), and the approximate distribution under the null is chi-square with degrees of freedom equal to the number of extra parameters in the larger model.

\subsection{Details}

Let $X_1,\ldots,X_{n_0} \in \bbZ_+$ and $X_{n_0+1},\ldots,X_{n_0+n_1} \in \bbZ_+$ denote observations from two groups. We know that each observation is from a mixed Poisson model $X_i \sim \Poisson(\theta_i)$, where $\theta_i \sim g_*^0(\cdot)$ if $1 \leq i \leq n_0$ and $\theta_i \sim g_*^1(\cdot)$ if $n_0 + 1 \leq i \leq n_0+n_1$.

Given a set of possible distributions given by $\cG$, we can estimate the mixing distributions via nonparametric MLE:
\[ \max_{(g^0,g^1) \in \cG} \left\{ \underbrace{\sum_{i \in [n_0]} \log\left( \bbE_{\theta \sim g^0} \left[ \frac{X_i^\theta \exp(-\theta)}{X_i!} \right] \right) + \sum_{i \in [n_1]} \log\left( \bbE_{\theta \sim g^1} \left[ \frac{X_{n_0+i}^\theta \exp(-\theta)}{X_{n_0+i}!} \right] \right)}_{:=\ell(g_0,g_1)} \right\}. \]

We use a \emph{likelihood ratio test} to test null hypothesis that $g^0$ and $g^1$ are the same. This test considers the ratio of the likelihood of the observations given freedom to choose $g^0$ and $g^1$ separately versus the likelihood with the constraint that $g^0 = g^1$. We know from our earlier discussion in \cref{sec:estimation} that the maximum likelihood occurs at a finite support distribution. For simplicity we assume that the log likelihood maximisations occur over the set of distributions with $K$ atoms, thus we optimize the log likelihood over this set.

We implemented this via a heuristic based on \cref{alg:CG}. First, we used a trust region method to optimize the log-likelihood over all distributions $g = \sum_{k=1}^{K_0} p_k \delta_{\lambda_k}$, i.e., we find parameters $(p_1,\ldots,p_{K_0},\lambda_1,\ldots,\lambda_{K_0})$ such that the log-likelihood is high. We then ran $K - K_0$ iterations of \cref{alg:CG} to generate two other points, so that we now have $K$ points. Finally, we used this $K$-point distribution as a starting point for another trust region search to improve the log likelihood further. We found that even with small values of $K$, this heuristic method was able to match, or closely match, likelihood values found by running \cref{alg:CG} to termination. We thus used $K_0=3$, $K=5$ in our experiments.


Our purpose here is to identify when there are different underlying distributions for the two groups and in practice a relatively small $K$ will already be optimal for the full maximum likelihood distribution. We tested on some examples and found that using the conditional gradient heuristic with $K=5$ achieved a similar optimal value as applying the full version of \cref{alg:CG} to the maximum likelihood problem. Thus, in our experiments, we used $K=5$.

We let $\cG_K$ be the set of distributions on $[0,\infty)$ with $K$ atoms. Then the two sets we consider are
\begin{align*}
	\cG_{\neq} &:= \cG_K \times \cG_K\\
	\cG_{=} &:= \left\{ (g^0,g^1) \in \cG_K \times \cG_K : g^0 = g^1 \right\},
\end{align*}
so $\cG_{=} \subset \cG_{\neq}$ as required for the LR test. We have a null hypothesis that $(g_*^0,g_*^1) \in \cG_{=}$, and define
\[ \left( \hat{g}^0, \hat{g}^1 \right) := \argmax_{(g^0,g^1) \in \cG_{\neq}} \ell(g_0,g_1), \quad \left( \hat{g}_{=}, \hat{g}_{=} \right) := \argmax_{(g^0,g^1) \in \cG_{=}} \ell(g_0,g_1).\]
The test statistic for the likelihood ratio test is
\[ \sigma_{\text{LR}} = -2 \left[ \ell\left( \hat{g}_{=}, \hat{g}_{=} \right) - \ell\left( \hat{g}^0, \hat{g}^1 \right) \right]. \]
Under the null hypothesis, $\sigma_{\text{LR}}$ converges asymptotically to the $\chi^2$-distribution, with degrees of freedom equal to the difference in dimension between $\cG_{\neq}$ and $\cG_{=}$, which is $2K-1$.

\subsection{Experiments}

Before giving details of the numerical results we state our main conclusion. Our approach of using a hypothesis test to determine whether or not to separate two groups is successful, but the improvement in final profits as a result of using this approach is very small. We can show a statistically significant improvement with the right choices, but the improvement in profits is not likely to be significant. Another way to state this is that in the tradeoff between either having a more accurate assessment of the underlying distribution from more samples, or making the right choice between two different possible distributions, the countervailing effects often cancel each other out.

We carry out a series of experiments in which we consider two groups of different sizes and having possibly different values of the Weibull scale parameter $\beta$. In each run the larger group has size $n_0$ taken from the set $\{1000,5000,10000\}$, while the smaller group has size $n_1$ given by 10\%, 20\% or 50\% of this.  Thus, there are 9 size pairings in total. The values of $\beta$ for the larger group ($\beta_0$) are drawn from $\{2,4,8\}$ and the smaller group has a $\beta$ value ($\beta_1$) given by a multiple $0.6$, $0.8$, $1$, $1.2$, $1.4$ or $1.6$
of the larger group $\beta$. As in the previous experiments, each unit $i$ has $r_i = 1$ and a cost $c_i$ drawn randomly from the range $[0.5,0.9]$. Different runs use different values for the base cost with $b_i = b \in \{0, 0.1,0.2,0.3,0.4,0.5,0.6\}$. For each set of parameters $n_0, n_1, \beta_0, \beta_1, b$ we make 30 runs, giving $3 \times 3 \times 3 \times 6 \times 7 \times 30 = 34,020$ runs in total.

We have carried out a series of experiments looking at whether or not it is better to consider a subset separately, using the $p$-value in a likelihood ratio test (with $K=5$) as the means of making that decision.  For simplicity we just consider $p$-values of 1, 0.5, 0.05, 0.005 etc. An example of the results from one of these experiments is shown in Table \ref{tab:grouping_1000x200}, which is for the case where the smaller group is of size 200 and the larger group is of size 1000. With these choices of $n_0$ and $n_1$ there are a total of 3780 runs with varying scale parameters for the Weibull, differing between the two subsets in most cases but including 630 runs in which the two subsets are drawn from the same underlying distribution.  For example, consider the row of the table with cutoff 0.05. This corresponds to a case where the LR test is carried out with this significance level and the smaller subset is treated on its own for any run in which the LR test rejects the hypothesis that the the two subsets are drawn from the same distribution. We show the average profit (over all runs) for the smaller group with this cutoff policy, and also the  proportion of the total set of runs for which this policy will consider the smaller group separately.

\begin{table} [hbt!]
	\centering
	\caption{Comparison of different $p$ value cutoffs ($n_0=200$, $n_1=1000$)} \label{tab:grouping_1000x200}
	\begin{tabular}{clll}
		\toprule
		p value & Average profit & Proportion of & Reduction in profit \\Cutoff & group 1 & runs split &  conditional on change \\
		\midrule
		0.0005 & \textbf{0.445217} & 0.452 & \\
		0.000005 & 0.445144 & 0.348 & 0.000707\quad (0.000590)\\
		0.00005 & 0.445120 & 0.404 & 0.002011\quad	(0.000917)\\
		0.005 & 0.445023 & 0.541 & 0.002188\quad (0.000730)\\
		0.05 & 0.444898 & 0.637 & 0.001722\quad (0.000412)\\
		0.5	 & 0.444085 & 0.811 & 0.003152\quad (0.000285)\\
		1    & 0.443236 & 1  & 0.003615\quad (0.000208)\\
		\bottomrule
	\end{tabular}
\end{table}

We can see in the table that the total profit is relatively insensitive to changes in the cutoff level, but is maximised in this set by taking a cutoff of 0.0005, which we have put in the first row. This implies treating the subset separately in about 45\% of runs. The small differences we observe in average profit can be significant since only a portion of the total runs switch from being grouped together or being separated when the $p$-value cutoff is changed. The final column gives the average change (for those runs where there is a change) when comparing with the best choice of cutoff in the top row, together with standard errors shown in brackets. We see that the differences between 0.0005 and lower values are not very significant here, but other differences, though small, are nevertheless statistically significant being at least as large as 3 standard errors.

The low value of $p$ here implies that with these group sizes we want to be very sure that there is a difference before deciding to treat the subset separately. This is because of the benefits of making better estimates through the larger combined set of observations.

However, as the sizes of the two groups increase, it becomes better to split even if we are not sure that the two groups are drawn from different distributions.  This is illustrated in Table \ref{tab:summary_grouping} where in each case we show the p value that gives the highest profit when used as a cutoff. The columns correspond to the larger group size (1000,5000 and 10000) and the rows to the smaller group set as 10\% , 20\% and 50\% of the larger group.

\begin{table} [hbt!]
	\centering
	\caption{Best choice of p for different combinations of group sizes}
	\label{tab:summary_grouping}
	\begin{tabular}{cccc}
		\toprule
		$n_0$ & $n_1=1000$ & $n_1=5000$ & $n_1=10000$\\
		\midrule
		100  & 0.5 &  & \\
		200  & 0.0005 &  & \\
		500  & 0.0005 & 0.005 & \\
		1000 &  & 0.05 & 0.5 \\
		2000 &  &  & 0.5\\
		2500 &  & 0.05 & \\
		5000 &  &  & 0.5\\
		\bottomrule
	\end{tabular}
\end{table}

We can see that in general increasing the sizes of the two groups leads to higher $p$-values at the best cutoff, implying that the two groups are treated separately for a greater proportion of the runs.  The reason for this is that the small group on its own already contains enough items to make a good empirical Bayes estimate with little gain from including more items.  Thus, it makes sense to split even if there is not much confidence that the two groups are different. However, in no case do we want to always split (i.e. take cutoff $p=1$).

There is one anomaly which is the 1000 by 100 case. Here the experiments suggest a high cutoff of $p=0.5$, rather than $p=0.0005$.  However there is no significant difference between this choice giving a profit of 0.453294, and using $p=0.0005$ which gives a profit of 0.449650.  In fact, the average difference for the runs where there is a change between the two cutoff values is 0.008131, with a standard error of 0.01230.

In all these experiments we have just investigated whether it is best to take the smaller group and consider it on its own. It is quite possible that in a particular run it is best to combine groups in determining the optimal policy for the smaller group, but better to split when determining the optimal policy for the larger group.  For that reason it is better not to assume that the same decision is made for both groups.

In our experiments we looked also at the larger group (which here is always of size at least 1000).  The differences between different cutoffs are often not significant and so there are some inconsistencies when we try to define a best choice. However, a good heuristic rule is that for the best profit for the larger group we should consider it on its own whenever the $p$-value is less than 0.5. 
\section{Conclusion}
One of the most fundamental inventory problems occurs when there are multiple items each experiencing Poisson demand, with a rate parameter $\lambda_i$ for item $i$, and we need to solve a newsvendor inventory problem for each item. A natural method for leveraging shared information across items is Empirical Bayes, yet this approach has received surprisingly little attention in the inventory management literature. In this paper we explore the value that can be gained from an empirical Bayes approach. Our focus has been on three practical questions which arise in this context.

We find that empirical Bayes can significantly improve decision quality, but the way it is applied matters. Either we may use using empirical Bayes to estimate the Poisson rate for each item and use that estimate to determine the optimal inventory policy, or we may use empirical Bayes to estimate for each item the probability of different levels of demand. We show that the second approach is preferred and through an asymptotic analysis we demonstrate that simply estimating the Poisson rate for each item may lead to highly suboptimal inventory decisions.

The paper also focuses on how to implement empirical Bayes effectively in practice. The full empirical Bayes approach involving an estimate of the probability of each potential demand level is computationally demanding. Incorrect choices in the way in which these calculations are carried out can dramatically lower the benefits to be derived from empirical Bayes. The recommended approach is to model the distribution of the rate parameter $\lambda_i$ as a mixture with finite support. This ($g$-modelling) approach is shown to be effective across a range of different environments. 

    Finally we look at the question of when we should partition items into different types before applying empirical Bayes. The advantage of doing this is to reduce the differences in demand patterns we see for different items leading to increased accuracy in our estimates, but at the same time this comes at the cost of smaller sample sizes within each group, making the empirical Bayes approach less effective.  Our experimental results indicate that the impact of choosing the right balance between these competing considerations is small. Typically, there will be very little improvement in profit that is possible through the decisions on whether or not to consider different types separately.

\bigskip

\bigskip

\begin{appendices}

\section{Supplement to \cref{Asymptotics}}\label{app:asymptotics}

\subsection{Proofs}\label{app:proofs}

\proof{Proof of \cref{half.norm.prop}.}
	Let
	\begin{equation*}
		I_k=\int_0^{\infty}\lambda^k e^{-\lambda}e^{-\frac{\lambda^2}{2}}d\lambda.
	\end{equation*}
	Note that $f(k)=\sqrt{2}I_k/(k!\sqrt{\pi})$, and hence
	\begin{equation*}
		R(k)=\frac{f(k+1)}{f(k)}=\left(\frac{1}{k+1}\right)\frac{I_{k+1}}{I_k}.
	\end{equation*}
	To complete the proof, we will establish $I_{k+1}/I_k\lesssim \sqrt{k\log(k)}$. Let $L_k=\sqrt{3(k+1)\log(k+1)}$ and note that
	\begin{equation*}
		I_{k+1}=\int\limits_0^{L_k}\lambda\lambda^k e^{-\lambda}e^{-\frac{\lambda^2}{2}}d\lambda
		+\int\limits_{L_k}^{\infty}\lambda^{k+1} e^{-\lambda}e^{-\frac{\lambda^2}{2}}d\lambda=A_1+A_2.
	\end{equation*}
	We have
	\[ A_1\le \int\limits_0^{L_k} L_k \lambda^k e^{-\lambda}e^{-\frac{\lambda^2}{2}}d\lambda \le L_k I_k. \]
	Notice that for $k \ge 2$, $1 \le 3 \log (k+1) \le k+1$, and so $L_k \le k+1$. Also, for $\lambda \ge L_k$, $(k+1) \log \lambda - \lambda ^2/2$ is decreasing in $\lambda$. Thus for $\lambda \ge L_k$,
	\begin{align*}
		(k+1)\log(\lambda) - \lambda ^2 /2 & \le  (k+1)\log(L_k) - (3/2) (k+1) \log(k+1) \\
		& \le  (k+1) (\log(k+1) - (3/2) \log(k+1)) < 0.
	\end{align*}
	Thus
	\begin{equation*}
		A_2=\int\limits_{L_k}^{\infty} \exp\big([k+1]\log(\lambda)-\lambda^2/2-\lambda\big)d\lambda\le \int\limits_{L_k}^{\infty} \exp(-\lambda)d\lambda<1.
	\end{equation*}
	Consequently, $I_{k+1}\lesssim \sqrt{k\log(k)}I_k$ and $R(k)\lesssim \sqrt{\log(k)/k}$.
\qed
\endproof

\proof{Proof of \cref{asymptotics.naive.thm}}
	Let~$h_*$ be a positive constant that satisfies $(1-1/h_*)^2>1-C_L$. Let $d_*=(1-1/h_*)^2 - 1+C_L$ and note that $d_*>0$. Define $S(x)=(x+1)R(x)$. From \cref{ass:Rk-small},  $S(k)=o(k)$ as $k\rightarrow\infty$, and hence
	\begin{equation}
		\label{ineq.S}
		\frac{S(k)}{k}\le \min\left\{\frac18\,,\,\frac{d_*}{4h_*^2(1+d_*)}\right\}
	\end{equation}
	will hold for all sufficiently large~$k$. We will let $b\ge B$, where $B$ is such that inequality~(\ref{ineq.S}) holds all $k\ge BC_L$.
	
	Take an arbitrary $i\in[n]$. Recall that, ignoring the per item cost~$b$, the na\"{i}ve allocation for item $i$ assumes a Poisson distribution for the future demand $X^{new}_{i}$ having rate parameter (and mean) given by the observation $X_i$. Then the na\"{i}ve allocation for item~$i$ is the $(1-c_i)$ quantile of this distribution.
	
	The Markov inequality implies that $\Poisson(X_i)$ is greater than $X_i/C_L$ with probability less than $C_L$. Hence the $1-C_L$ quantile is no more than $X_i/C_L$.  Because $c_i\ge C_L$, the na\"{i}ve allocation for item~$i$ after observing $X_i$ is at most $X_i/C_L$. The revenue per item is normalised to 1 so the corresponding profit is at most $X_i/C_L - b$. Hence $b>X_i/C_L$ implies from (\ref{condition x_i=0}) that the na\"{i}ve allocation is zero. Consequently, to understand the profit of the na\"{i}ve allocation it is sufficient to focus on the case $X_i\ge bC_L$. Note that in this case, inequality~(\ref{ineq.S}) implies
	\begin{equation}
		\label{ineq.Xi}
		\frac{S(X_i)}{X_i}\le \min\left\{\frac{1}{8}\,,\,\frac{d_*}{4h_*^2(1+d_*)}\right\}.
	\end{equation}
	
	The Robbins formula gives $\bbE [\lambda_{i} \mid X_{i}]=S(X_i)$. From this we can use the Markov inequality to show that $\bbP \Big[\lambda_{i}\le h_*S(X_i) \mid X_{i}\Big]\ge 1-1/h_*$. The Markov inequality also gives $\bbP \big[\Poisson(\lambda)\le h_*\lambda) \big] \ge 1-1/h_*$. Using these two inequalities, we can derive
	\begin{align*}
		\label{ineq.istar}
		\bbP \Big[X^{new}_{i}\le h_*^2S(X_i) \mid X_{i}\Big] & \ge \bbP \Big[\lambda_{i}\le h_*S(X_i) \mid X_{i}\Big] \; \bbP\Big[ X^{new}_{i}\le h_*^2S(X_i) \mid X_{i},\lambda_{i}\le h_*S(X_i)\Big] \\
		&\ge (1-1/h_*) \; \bbP \Big[\Poisson(\lambda_i)\le h_*^2S(X_i) \mid X_{i},\lambda_{i} = h_*S(X_i)\Big] \\
		&\ge (1-1/h_*)^2 > 1-C_L.
	\end{align*}
	For the second inequality above we have used the fact that $X_i^{new}$ is a mixture of $\Poisson(\lambda)$ for $\lambda$ in an interval, and so the left-tail probability for this distribution will be minimized at the largest value of~$\lambda$.
	
	Because $c_i\ge C_L$, the optimal allocation for item~$i$ will have at most~$h_*^2S(X_i)$ units. We will show that the na\"{i}ve allocation will sometimes allocate more and lose money as a result.
	
	The inequality above gives
	\begin{equation}
		\label{prob.bnd.tail}
		\bbP \Big[ X^{new}_i> h_*^2S(X_i) \mid X_i \Big] \le 1-(1-1/h_*)^2 =C_L-d_*.
	\end{equation}
	Let~$A_{i}$ denote the na\"{i}ve allocation for item~$i$, and let $K_i=A_i-\lceil h_*^2S(X_i)\rceil$. When $K_i>0$ bound~(\ref{prob.bnd.tail}) implies that the (conditional) expected revenue for each of the last $K_i$ units in the na\"{i}ve allocation for item~$i$ is bounded above by $C_L-d_*$. Because~$c_i\ge C_L$, the combined expected profit for these~$K_i$ units is bounded above by $-d_*K_i$.
	At the same time, even ignoring all the costs, the combined expected profit for the first $\lceil h_*^2S(X_i)\rceil$ units in the na\"{i}ve allocation for item~$i$ is bounded above by $\lceil h_*^2S(X_i)\rceil$. Consequently, the expected profit of the na\"{i}ve allocation for item~$i$ is bounded above by $-d_*K_i+\lceil h_*^2S(X_i)\rceil$.
	
	Using Jensen's inequality, we derive
	\begin{equation}
		\label{zero.prob}
		\bbP [X^{new}_i=0 \mid X_i]=\bbE [e^{-\lambda_i} \mid X_i] \ge e^{-\bbE [\lambda_i \mid X_i]} = e^{-S(X_i)}.
	\end{equation}
	The inequality $\bbP [\Poisson(\lambda)\le (1-\epsilon)\lambda ] \le e^{-\epsilon^2\lambda/2}$, holds for every positive~$\lambda$ and~$\epsilon$ as is shown by \citet{alon2016probabilistic}. This implies that
	\begin{equation}
		\label{prob.bnd.half}
		\bbP \Big[\Poisson(X_i)\le X_i/2 \mid  X_i\Big]\le e^{-X_i/8}.
	\end{equation}
	If $1-c_i\le e^{-X_i/8}$, then $1-c_i\le e^{-S(X_i)}$ by~(\ref{ineq.Xi}), and the optimal allocation is zero by~(\ref{zero.prob}), which implies that the corresponding expected profit for the na\"{i}ve allocation cannot be positive. On the other hand, if $1-c_i> e^{-X_i/8}$, then the na\"{i}ve allocation is at least~$X_i/2$ by~(\ref{prob.bnd.half}).
	
	It follows that on the event $\mathcal{E}_i=\{X_i\ge bC_L, 1-c_i> e^{-X_i/8}\}$, the total expected profit for the na\"{i}ve allocation for item~$i$ is bounded above by $ -d_*[X_i/2-\lceil h_*^2S(X_i)\rceil]+\lceil h_*^2S(X_i)\rceil$, which is at most $-X_id_*/4$ by inequality~(\ref{ineq.Xi}). Outside of the event $\mathcal{E}_i$, the total expected profit for the na\"{i}ve allocation for item~$i$ is non-positive. Summing over~$i$, we conclude that the total expected profit for the na\"{i}ve allocation satisfies:
	\begin{equation*}
		\widehat{\Pi}_n \le \sum_{i=1}^n  \Big(-\frac{X_id_*}{4}\Big)\mathbf{1}_{\mathcal{E}_i}\le -n \left(\frac{bC_Ld_*}{4}\right)  \left(\frac1n\sum_{i=1}^n  \mathbf{1}_{\mathcal{E}_i}\right).
	\end{equation*}
	Note that $\bbP [ \mathcal{E}_i ]>0$ by the assumptions on the distributions of~$X_i$ and~$c_i$. Hence, by the law of large numbers, $\widehat{\Pi}_n \lesssim -n$ with probability tending to one.
\qed
\endproof

\proof{Proof of \cref{asymptotics.plugin.thm}.}	
	Suppose that $T=1$, $b=0$, $r_i=1$, and~$c_i=c$ for all~$i$. For most of the proof, we will focus on an arbitrary item~$i$ but omit the corresponding subscript~$i$ from the notation to simplify the presentation. For example, we will write~$X$ for the observed demand on this item, and write~$\lambda$ for the rate of the underlying Poisson distribution. Let the support of the prior for~$\lambda$ be a finite set of~$K$ values $\lambda_1>\lambda_2>...>\lambda_K$, where $K>1$, and denote $\bbP [\lambda=\lambda_j]$ by~$p_j$ for~$j=1,...,K$, where all the~$p_j$ are positive.
	
	Note that
	\begin{equation*}
		\bbP [\lambda=\lambda_k \mid X=s] =\frac{\lambda_k^s e^{-\lambda_k}p_k}{\sum_{j=1}^K\lambda_j^s e^{-\lambda_j}p_j} \qquad\text{and}\qquad \bbP [X^{\text{new}}=0 \mid X=s] = \frac{\sum_{k=1}^K\lambda_k^{s} e^{-2\lambda_k}p_k}{\sum_{j=1}^K\lambda_j^s e^{-\lambda_j}p_j}.
	\end{equation*}
	Define $q=\lambda_2/\lambda_1$ and note that $q<1$. As $s\rightarrow\infty$, we have
	\begin{equation}
		\label{p.zero.true}
		\bbP [X^{\text{new}}=0 \mid X=s] = \frac{e^{-2\lambda_1}p_1 + e^{-2\lambda_2}p_2q^s + o(q^s)}{e^{-\lambda_1}p_1 + e^{-\lambda_2}p_2q^s + o(q^s)}=e^{-\lambda_1}+a_*[1+o(1)]q^s,
	\end{equation}
	where~$a_*=\big( e^{\lambda_1-\lambda_2}-1 \big) \big(e^{-\lambda_2}p_2/p_1\big)$.
	
	We also have
	\begin{equation*}
		\bbE [\lambda \mid X=s] = \frac{\sum_{k=1}^K\lambda_k^{s+1} e^{-\lambda_k}p_k}{\sum_{j=1}^K\lambda_j^s e^{-\lambda_j}p_j}=\lambda_1-e^{\lambda_1-\lambda_2}(\lambda_1-\lambda_2)(p_2/p_1)q^s+o(q^s).
	\end{equation*}
	Writing $\tilde{P}$ for the (incorrect) probability according to the plugin approach, we can derive
	\begin{equation}
		\label{p.zero.plugin}
		\tilde{P}(X^{\text{new}}=0 \mid X=s)=e^{-\bbE [\lambda \mid X=s]} = e^{-\lambda_1}+\tilde{a}[1+o(1)]q^s,
	\end{equation}
	where~$\tilde{a}=\big(\lambda_1-\lambda_2\big)\big(e^{-\lambda_2}p_2/p_1\big)$. We also have
	\begin{equation}
		\label{p.one.plugin}
		\tilde{P}(X^{\text{new}}=1 \mid X=s) = \bbE [\lambda \mid X=s] \; e^{-\bbE[\lambda \mid X=s]} = \lambda_1 e^{-\lambda_1}+O(q^s).
	\end{equation}
	
	Note that $e^{\lambda_1-\lambda_2}-1 > \lambda_1-\lambda_2 > 0$, and hence $a^*>\tilde{a}>0$. We will take~$a$ as a fixed constant in $(\tilde{a},a_*)$ and let
	\begin{equation}
		\label{c.def}
		c:=1-e^{-\lambda_1}-a q^S
	\end{equation}
	for some large positive integer~$S$ that will be specified later.

	Equations~(\ref{p.zero.true})-(\ref{c.def}) imply that by making~$S$ sufficiently large, we can ensure that the following inequalities hold for some positive constant~$\delta$:
	\begin{eqnarray}
		\bbP [X^{\text{new}}=0 \mid X=S]&\ge& 1-c + \delta q^S \label{P.0.eq}\\
		\nonumber\\
		\tilde{P}(X^{\text{new}}=0 \mid X=s)&<&1-c, \qquad\;\; \text{for}\; s\ge S \label{Ptilde.0.eq}\\
		\nonumber\\
		\tilde{P}(X^{\text{new}}\le 1 \mid X=s)&>& 1-c, \qquad\;\; \text{for}\; s\ge S. \label{Ptilde.01.eq}
	\end{eqnarray}
	By~(\ref{Ptilde.0.eq}) and~(\ref{Ptilde.01.eq}), when the observed $X$ is at least~$S$, the plugin method will make an allocation of exactly one unit. Thus, the corresponding (conditional) expected profit is $\bbP [X^{\text{new}}>0 \mid X=s]-c$, which can be bounded above by $1-c- \bbP [X^{\text{new}}=0 \mid \lambda=\lambda_1]=aq^S$ using the definition of~$c$ in~(\ref{c.def}) and the fact that $\lambda\le\lambda_1$. Moreover, when~$X=S$ the corresponding expected profit is at most $-\delta q^S$ by~(\ref{P.0.eq}). Also note that when~$X<S$, the probability that $X^{\text{new}}=0$ is larger than the right hand side of (\ref{P.0.eq}) using \cref{P0.monot.lem}. Thus the optimal allocation is zero with zero profit, and hence the corresponding expected profit for the plugin approach cannot be positive.
	
	Consequently,
	\begin{equation}
		\label{plugin.profit.expr}
		\Pi_n^{\text{plug}}\le -n q^S \Big(  \frac{\delta}n\sum_{i=1}^n\mathbf{1}_{\{X_i=S\}} - \frac an\sum_{i=1}^n\mathbf{1}_{\{X_i>S\}}\Big),
	\end{equation}
	Note that $\bbP [X_i=S]>0$ and
	\begin{equation*}
		\frac{\bbP [X_i>S]}{\bbP [X_i=S]}=\sum_{k=1}^{\infty}\frac{\bbP [X_i=S+k]}{\bbP [X_i=S]}
		\le \sum_{k=1}^{\infty}\frac{\lambda_1^{k}}{(S+1)\cdots(S+k)} \le \sum_{k=1}^{\infty}\Big(\frac{\lambda_1}{S+1}\Big)^k = \frac{\lambda_1}{S+1-\lambda_1}.
	\end{equation*}
	Thus, by making~$S$ sufficiently large, we can ensure that $\delta \bbP [X_i=S] - a \bbP [X_i>S] > (\delta/2) \bbP [X_i=S]$, which implies that $\Pi_n^{\text{plug}}\lesssim -n$ by~(\ref{plugin.profit.expr}) and the law of large numbers.
	
	Now observe from (\ref{p.zero.true}) that $\bbP [X^{\text{new}}=0 \mid X=s] < 1-c$ for all sufficiently large~$s$ (so that the term in $q^s$ is dominated by the term in $q^S$). Thus for the optimal policy there will be a positive profit for items where the observed demand is high enough. This is enough to establish the second part of the theorem that $\Pi_n^* \gtrsim n$.
\qed
\endproof

\begin{lemma}\label{P0.monot.lem}
	Under the setting of Theorem~\ref{asymptotics.plugin.thm}, $\bbP [X^{\text{\rm new}}=0 \mid X=k]$ is a strictly decreasing function of~$k$.
\end{lemma}
\proof{Proof of \cref{P0.monot.lem}.}	
	Take an arbitrary non-negative integer~$k$. We will show that $\bbP [X^{\text{new}}=0 \mid X=k] = \bbE [e^{-\lambda} \mid X=k]$ is greater than $\bbP [X^{\text{new}}=0 \mid X=k+1] = \bbE [e^{-\lambda} \mid X=k+1]$.
	
	To simplify the expressions, we will denote the support of~$\lambda$ by $\tilde{\lambda}_1<\tilde{\lambda}_2<...<\tilde{\lambda}_K$ and write $p(j \mid k)$ for the conditional probability $\bbP [\lambda=\tilde{\lambda}_{j} \mid X=k]$. First, note that for every $j\in\{1,...,K-1\}$,
	\begin{equation}
		\label{ratio.rel}
		\frac{p(j+1 \mid k)}{p(j \mid k)} < \frac{p(j+1 \mid k+1)}{p(j \mid k+1)}
	\end{equation}
	since the right hand side is equal to the left hand side multiplied by $\tilde{\lambda}_{j+1}/\tilde{\lambda}_{j}$. Informally, this means that the relative conditional likelihood of larger values of~$\lambda$ is greater when $X=k+1$ than when $X=k$. We will show that this leads to a smaller conditional mean of~$e^{-\lambda}$ when $X=k+1$.
	
	Denote by $E_{k,m}$ the conditional expectation given $X=k,\lambda\in\{\tilde{\lambda}_1,...,\tilde{\lambda}_m\}$. When $m=2$, we have $p(1 \mid k)+p(2 \mid k)=1$ and inequality~(\ref{ratio.rel}) with~$j=1$ implies that
	\begin{equation}
		\label{Ek2.ineq}
		E_{k,2}(e^{-\lambda}) > E_{k+1,2}(e^{-\lambda}).
	\end{equation}
	We also have
	\begin{equation}
		\label{Ek3.eq}
		E_{k,3}(e^{-\lambda}) = E_{k,2}(e^{-\lambda})\left[\frac{p(1 \mid k)+p(2 \mid k)}{p(1 \mid k)+p(2 \mid k)+p(3 \mid k)}\right]
		+e^{-\tilde{\lambda}_3}\left[\frac{p(3 \mid k)}{p(1 \mid k)+p(2 \mid k)+p(3 \mid k)}\right].
	\end{equation}
	By inequality~(\ref{ratio.rel}),
	\begin{equation}
		\label{p3k.ineq}
		\frac{p(3 \mid k)}{p(1 \mid k)+p(2 \mid k)+p(3 \mid k)}<\frac{p(3 \mid k+1)}{p(1 \mid k+1)+p(2 \mid k+1)+p(3 \mid k+1)},
	\end{equation}
	i.e., the second weight in the weighted average~(\ref{Ek3.eq}) is smaller than the one in the corresponding expression for $E_{k+1,3}(e^{-\lambda})$. Inequalities $e^{-\tilde{\lambda}_3}< E_{k,2}(e^{-\lambda})$,  (\ref{Ek3.eq}), and~(\ref{p3k.ineq}) imply that
	\begin{align*}
		E_{k,3}(e^{-\lambda}) &> E_{k,2}(e^{-\lambda})\left[\frac{p(1 \mid k+1)+p(2 \mid k+1)}{p(1 \mid k+1)+p(2 \mid k+1)+p(3 \mid k+1)}\right]\\
		&\quad +e^{-\tilde{\lambda}_3}\left[\frac{p(3 \mid k+1)}{p(1 \mid k+1)+p(2 \mid k+1)+p(3 \mid k+1)}\right].
	\end{align*}
	Applying inequality~(\ref{Ek2.ineq}) to the first term on the right-hand side, we arrive at $E_{k,3}(e^{-\lambda}) > E_{k+1,3}(e^{-\lambda})$.
	Continuing by induction, we conclude that $E_{k,K}(e^{-\lambda}) > E_{k+1,K}(e^{-\lambda})$, which is equivalent to $\bbP [X^{\text{new}}=0 \mid X=k] > \bbP [X^{\text{new}}=0 \mid X=k+1]$.
\qed
\endproof

\subsection{Further properties of~$\Pi_n^*$}\label{app:opt.profit}

First, we show that when the (prior) distribution of~$\lambda$ is unbounded, $\Pi_n^*\gtrsim n$ with probability tending to one as $n$ tends to infinity. As before, we will focus on a arbitrary item~$i$ and drop the subscript~$i$ from the notation to simplify the presentation.

We start by focusing on the case where $c= C$ for some fixed constant~$C<1$ and then generalize to the case of random~$c$. Let~$x^*_k$ denote the optimal allocation given that $X=k$ is observed, and recall that $x^*_k$ is the $(1-C)$-quantile of the distribution of~$X^{\text{new}}$. For every fixed constant~$M$, we have
$\bbP[\lambda>M|X=k]\rightarrow1$ as $k\rightarrow\infty$, which implies $x^*_k\rightarrow\infty$ as $k\rightarrow\infty$. Note also that there exists a positive constant~$\delta$, such that, for all sufficiently large~$k$, the conditional expected profit for the item under consideration is at least $\delta x^*_k(1-C) -b$. Consequently, we can find a~$k$ large enough, so that this value is positive. Because $\bbP[X=k]>0$, the law of large numbers then implies that the total expected profit satisfies $\Pi_n^*\gtrsim n$ with probabiltity tending to one. The general case of random~$c$ follows from the observation that there exists a constant $C<1$ such that  $\bbP[c\le C]>0$.

Second, we consider the case where the distribution of~$\lambda$ is bounded. Suppose, for concreteness, that $c=C$ for some constant~$C<1$, and let $\Lambda$ be the upper bound on~$\lambda$ such that $\bbP[\lambda > \Lambda-\epsilon]>0$ for every positive~$\epsilon$. Extending the arguments above, we note that $\Pi_n^*\gtrsim n$ with probability tending to one if the conditional expected profit (for one item) is positive under the condition that $\lambda=\Lambda$. In other words, we need
$b < \sum_{k=0}^{x^*-1} k p(k)$, where~$x^*$ is the $(1-C)$-quantile of the $\text{Poisson}(\Lambda)$ distribution, and $p(k)$ are the corresponding $\text{Poisson}(\Lambda)$ probabilities.

\section{Supplement to \cref{sec:estimation}}\label{app:estimation}

\subsection{Further details on \cref{sec:estimate-f}}

We will fit the following parametric model:
\[ f_{\beta,T}(s) = \exp\left( \beta_0 + w(s)^\top \beta \right). \]
Here $w(s)$ is a vector of basis functions so that $w(s)^\top \beta$ is a cubic spline, specifically:
\begin{align*}
	w(s) &= (s,s^2,s^3,(s-c_1)_+^3, \ldots,(s-c_m)_+^3)
\end{align*}
for chosen knot points $0 < c_1 < \ldots < c_m = \tau$. We impose some additional constraints on $\beta$ based on our problem structure.  Because these constraints may all be modelled via the exponential cone, and the objective also has this form, the problem we end up with is a conic program and off-the-shelf solvers such as Mosek may then be used to solve for $\beta_0,\beta$.

\paragraph{Natural spline constraints.}
Following \citet{efron2011tweedie}, we will fit a \emph{natural spline}, which requires that $w(s)^\top \beta$ is linear in $s$ for $s \geq \tau$.
Notice that when $s \geq \tau$, we have
\begin{align*}
	w(s)^\top \beta &= \beta_1 s + \beta_2 s^2 + \beta_3 s^3 + \sum_{j \in [m]} \beta_{3+j} (s-c_j)^3\\
	&= \left( \beta_3 + \sum_{j \in [m]} \beta_{3+j} \right) s^3 + \left( \beta_2 - 3 \sum_{j \in [m]} c_j \beta_{3+j} \right) s^2 + \left( \beta_1 + 3 \sum_{j \in [m]} c_j^2 \beta_{3+j} \right) s - \sum_{j \in [m]} c_j^3 \beta_{3+j}.
\end{align*}
Denote the coefficients for each monomial in $s$ via $a_0,a_1,a_2,a_3$, where the $a$ vectors are such that
\begin{align*}
	a_3^\top \beta &= \beta_3 + \sum_{j \in [m]} \beta_{3+j}\\
	a_2^\top \beta &= \beta_2 - 3 \sum_{j \in [m]} c_j \beta_{3+j}\\
	a_1^\top \beta &= \beta_1 + 3 \sum_{j \in [m]} c_j^2 \beta_{3+j}\\
	a_0^\top \beta &= - \sum_{j \in [m]} c_j^3 \beta_{3+j}.
\end{align*}
To make $\beta_0 + \beta^\top w(s)$ linear for $s \geq t_{m-3}$, we need two linear constraints:
\begin{align*}
	0 &= a_2^\top \beta\\
	0 &= a_3^\top \beta.
\end{align*}

\paragraph{Probability constraint.} Since $f_{\beta,T}$ is a probability mass function, we need $\sum_{s=0}^\infty f_{\beta,T}(s) = 1$. Notice that for $s \geq s_0 := \lceil \tau \rceil$, we get
\begin{align*}
	f_{\beta,T}(s) &= \exp\left( \beta_0 + a_0^\top \beta + \left( a_1^\top \beta \right) s \right)\\
	&= \exp\left( \beta_0 + a_0^\top \beta + \left( a_1^\top \beta \right) s_0 \right) \exp\left( a_1^\top \beta\right)^{s-s_0}.
\end{align*}
Therefore the probability constraint is
\begin{align*}
	1 &= \sum_{s=0}^{s_0-1} \exp\left( \beta_0 + w(s)^\top \beta \right) + \exp\left( \beta_0 + a_0^\top \beta + \left( a_1^\top \beta \right) s_0 \right) \sum_{s=s_0}^\infty \exp\left( a_1^\top \beta\right)^{s-s_0}\\
	&= \sum_{s=0}^{s_0-1} \exp\left( \beta_0 + w(s)^\top \beta \right) + \frac{\exp\left( \beta_0 + a_0^\top \beta + \left( a_1^\top \beta \right) s_0 \right)}{1 - \exp\left( a_1^\top \beta\right)}.
\end{align*}
We introduce another variable $\gamma$ and impose the two constraints
\begin{align*}
	1 &\geq \sum_{s=0}^{s_0-1} \exp\left( \beta_0 + w(s)^\top \beta \right) + \exp(\gamma),\\
	\exp(\gamma) &\geq \frac{\exp\left( \beta_0 + a_0^\top \beta + \left( a_1^\top \beta \right) s_0 \right)}{1 - \exp\left( a_1^\top \beta\right)},
\end{align*}
where the second of these constraints can also be written
\[
1 \geq \exp\left( \beta_0 + a_0^\top \beta + \left( a_1^\top \beta \right) s_0  - \gamma \right) + \exp\left( a_1^\top \beta \right).
\]
This guarantees that
\[ \sum_{s=0}^{\infty} f_{\beta,T}(s) = \sum_{s=0}^{s_0-1} f_{\beta,T}(s) + \sum_{s=s_0}^\infty f_{\beta,T}(s) \leq 1. \]


\paragraph{Monotonicity constraints.} Another constraint that we impose is related to a known property of the posterior means: $\bbE[\lambda \mid X=s+1] \geq \bbE[\lambda \mid X=s]$ for all $s$. Through Robbins' formula, this implies that
\[ \frac{(s+2) f_T(s+2)}{T f_T(s+1)} \geq \frac{(s+1) f_T(s+1)}{T f_T(s)}
\]
and hence
\[ \frac{s+2}{s+1} \geq \frac{f_T(s+1)^2}{f_T(s+2) f_T(s)}.
\]
Thus we have the following set of constraints on the parameters:
\[ \log\left( \frac{s+2}{s+1} \right) \geq (2 w(s+1) - w(s+2) - w(s))^\top \beta, \quad  s=0,1,2, \ldots, s_0 -2. \]

\subsection{Further details on \cref{sec:estimate-g}}



We describe the CG method to solve \eqref{eq:non-parametric-finitedim}. \citet{Simar1976} also provides a technique to adjust distributions $g_t$ if needed if they have more points in the support than the bound on $r$ established above; we ignore this step as it involves solving a moment problem.

Recall that the objective function and its gradient are
\[ h(v) = \sum_{s \in \cS} \frac{y_s}{n} \log(v(s)), \quad \grad h(v) = \left\{ \frac{y_s}{n v(s)} \right\}_{s \in \cS}. \]
Since $h$ is concave in $v$, the optimality condition for \eqref{eq:non-parametric-finitedim} at $v$ is $\grad h(v)^\top (v - \tilde{w}) \geq 0$ for all $\tilde{w} \in \cV$, or equivalently
\[ \sum_{s \in \cS} \frac{y_s \tilde{w}(s)}{n v(s)} \leq \sum_{s \in \cS} \frac{y_s}{n} = 1. \]
Furthermore note that
\[ \max_{\tilde{w} \in \cV} h(\tilde{w}) - h(v) \leq \max_{\tilde{w} \in \cV} \grad h(v)^\top (\tilde{w} - v) = \max_{\tilde{w} \in \cV} \sum_{s \in \cS} \frac{y_s \tilde{w}(s)}{n v(s)} - 1,\]
thus $\max_{\tilde{w} \in \cV} \sum_{s \in \cS} \frac{y_s \tilde{w}(s)}{n v(s)}$ may be used to bound the optimality gap of a solution $v$. Given $v$, computing $\max_{\tilde{w} \in \cV} \sum_{s \in \cS} \frac{y_s \tilde{w}(s)}{v(s)}$ is a linear optimization problem over $\cV$, thus the optimal solution is an extreme point of $\cV$. The extreme points of $\cV$ are correspond to measures $\tilde{g}$ concentrated on a single point, thus they are within the following set:
\[ \left\{ \left\{ \lambda^s \exp(-\lambda) \right\}_{s \in \cS} : \lambda \in \bbR_+ \right\} \subset \cV. \]
Therefore
\[ \max_{\tilde{w} \in \cV} \sum_{s \in \cS} \frac{y_s \tilde{w}(s)}{v(s)} = \max_{\lambda \geq 0} \sum_{s \in \cS} \frac{y_s}{v(s)} \lambda^s \exp(-\lambda). \]
The CG method solves this subproblem each iteration, but the objective is a sum of quasiconcave functions on $\bbR_+$, which in general may be nonconcave. Furthermore, it is univariate thus there exists efficient line search methods that may be employed. In fact, the objective is the ratio of convex functions, so a branch-and-bound procedure exists to find the global minimum \citep{Benson2006}. Also, each term $\lambda^s \exp(-\lambda)$ increases for $\lambda \leq s$ and decreases for $\lambda \geq s$, so the objective $\sum_{s \in \cS} \frac{y_s}{v(s)} \lambda^s \exp(-\lambda)$ cannot have a maximizer larger than the largest observation $s^* = \max_{s \in \cS} s$. The CG method to solve \eqref{eq:non-parametric-finitedim} is presented in \cref{alg:CG}.
\begin{algorithm}[ht]\caption{Fully corrective CG method for solving \eqref{eq:non-parametric-finitedim}.}\label{alg:CG}
	\KwData{Initial support $\Lambda_1 = \left\{\lambda_1,\ldots,\lambda_L\right\}$. Iteration bound $\cT$, stopping tolerance $\epsilon > 0$.}
	\KwResult{Final point $\bar{v} \in \cV$ and corresponding distribution $\bar{g}$.}
	\For{$t = 1,\ldots,\cT$}{
		Solve
		\[ \bar{p}_t = \argmax_{p=\{p_{\lambda}\} _{\lambda \in \Lambda_t}  }  \left\{ \sum_{s \in \cS} \frac{y_s}{n} \log\left( \sum_{\lambda \in \Lambda_t} p_{\lambda} \lambda^s \exp(-\lambda) \right) : p \geq 0, \sum_{\lambda \in \Lambda_t} p_\lambda = 1 \right\}. \]
		Set
		\[ \bar{v}_t = \left\{ \sum_{\lambda \in \Lambda_t} \bar{p}_{t,\lambda} \lambda^s \exp(-\lambda) \right\}_{s \in \cS} \in \cV. \]
		Compute
		\[ \bar{\epsilon}_t = \max_{\lambda \geq 0} \sum_{s \in \cS} \frac{y_s}{n \bar{v}_t(s)} \lambda^s \exp(-\lambda) - 1\]
		with maximizer $\bar{\lambda}_t$\;
		\If{$\bar{\epsilon}_t \leq \epsilon$}{STOP\;}
		Set $\Lambda_{t+1} = \Lambda_t \cup \{\bar{\lambda}_t\}$.
	}
	Return $\bar{v} = \bar{v}_t$ and $\bar{g} = \bar{g}_t = \sum_{\lambda \in \Lambda_t} \bar{p}_{t,\lambda} \delta_{\lambda}$.
\end{algorithm}

If we ever encounter $\bar{\epsilon}_t \leq \epsilon$ in \cref{alg:CG} then the returned $\bar{v}$ and $\bar{g}$ are $\epsilon$-optimal for \eqref{eq:non-parametric-finitedim} and \eqref{eq:non-parametric}. Adapting the analysis of \citet[Sec. 5.1]{JagabathulaEtAl2020}, we now show that \cref{alg:CG} is \emph{convergent}, i.e., there exists a sufficiently large finite iteration limit $\cT$ for which \cref{alg:CG} is guaranteed to return $\epsilon$-optimal solutions, and in fact $\cT = O(1/\epsilon)$. This relies on establishing a lower bound on the minimum eigenvalue of the Hessian of the objective over $v \in \cV$. The Hessian $\grad^2 h(v)$ is simply a diagonal matrix with entries $-y_s/v(s)^2$. However, there exists vectors in $v$ for which some entries are arbitrarily close to $0$, thus the minimum eigenvalue is unbounded below. Following the strategy of \citet[Sec. 5.1]{JagabathulaEtAl2020}, we first show that each entry of the iterates $\bar{v}_t$ in \cref{alg:CG} is bounded below by a constant $\gamma > 0$.

\begin{theorem}\label{thm:CG-iterate-bound}
	For $u > 0$ define
	\[ G_s(u) = \max_v \left\{ h(v) : v \in \cV, v(s) \leq u \right\}, \quad G(u) = \max_{s \in \cS} G_s(u) \]
	and
	\[ \gamma = \inf_u \left\{ u : u > 0, G(u) \geq h(\bar{v}_1) \right\}, \]
	where $\bar{v}_1 \in \cV$ is the first iterate of \cref{alg:CG}. Then $\gamma > 0$ and for each $s \in \cS$ and iterate $\bar{v}_t$ computed by \cref{alg:CG}, we have $\bar{v}_t(s) \geq \gamma$. Furthermore, a positive lower bound for $\gamma$ can be computed as follows:
	\[ \gamma \geq \min_{s \in \cS} \left\{ \bar{v}_1(s) \prod_{s' \in \cS \setminus \{s\}} \left( \frac{\bar{v}_1(s')}{(s')^{s'} \exp(-s')} \right)^{y_{s'}/y_s} \right\}. \]
\end{theorem}
\proof{Proof of \cref{thm:CG-iterate-bound}.}
	Notice that $G_s(u) \to -\infty$ as $u \to 0$, is concave in $u$ hence continuous, and is non-decreasing. Thus also $G(u) \to -\infty$ as $u \to 0$ and is continuous. Therefore $\gamma > 0$, and $G(\gamma) = h(\bar{v}_1)$. Furthermore, note that every iterate $\bar{v}_t$ can only increase the objective $h(\bar{v}_t)$, i.e., $h(\bar{v}_t) \geq h(\bar{v}_1)$. Now suppose that $\bar{v}_t(s) < \gamma$. Then from the definition of $\gamma$, we must have $G(\bar{v}_t(s)) < h(\bar{v}_1)$. Thus $G_s(\bar{v}_t(s))< h(\bar{v}_1)$, but also from the definition of $G_s$ we have $G_s(\bar{v}_t(s)) \ge  h(\bar{v}_t)$. This contradicts $h(\bar{v}_t) \geq h(\bar{v}_1)$, and so we must have  $\bar{v}_t(s) \ge \gamma$ as required.
	
	For the lower bound, notice that $\lambda^s \exp(-\lambda) \leq s^s \exp(-s)$ for each $s \in \cS$, since the left hand side achieves its maximum at $\lambda = s$. This means that $v(s) \leq s^s \exp(-s)$ for every $v \in \cV$. An upper bound on $G_s(u)$ can be constructed by replacing each entry $v(s')$ with $(s')^{s'} \exp(-s')$, except for $v(s)$, which we replace with $u$:
	\[ G_s(u) \leq \frac{y_s}{n} \log(u) + \sum_{s' \in \cS \setminus \{s\}} \frac{y_{s'}}{n} (s'-1) \log(s'). \]
	The $u$ that makes the right-hand side equal to $h(\bar{v}_1)$ is
	\[ \exp\left( \frac{n}{y_s} h(\bar{v}_1) - \sum_{s' \in \cS \setminus \{s\}} \frac{y_{s'}}{y_s} (s'-1) \log(s') \right) = \bar{v}_1(s) \prod_{s' \in \cS \setminus \{s\}} \left( \frac{\bar{v}_1(s')}{(s')^{s'} \exp(-s')} \right)^{y_{s'}/y_s}. \]
	Thus we have $\gamma \geq \min_{s \in \cS} \left\{ \bar{v}_1(s) \prod_{s' \in \cS \setminus \{s\}} \left( \frac{\bar{v}_1(s')}{(s')^{s'} \exp(-s')} \right)^{y_{s'}/y_s} \right\}$.
\qed
\endproof

Equipped with \cref{thm:CG-iterate-bound}, we now provide the convergence rate of \cref{alg:CG}. Note that this is not as simple as replacing the domain $\cV$ with $\cV \cap \{ v : v \geq \gamma \bm{1} \}$. The reason is because, as discussed above, the subproblems for $\bar{\epsilon}_t$ we solve correspond to a linear optimization over $\cV$:
\[ \max_{\lambda \geq 0} \sum_{s \in \cS} \frac{y_s}{n \bar{v}_t(s)} \lambda^s \exp(-\lambda) = \max_{v \in \cV} \sum_{s \in \cS} \frac{y_s v(s)}{n \bar{v}_t(s)}. \]
This is not equivalent to solving over $\cV \cap \{ v : v \geq \gamma \bm{1} \}$, and imposing the constraint $v \geq \gamma \bm{1}$ complicates things greatly. Nevertheless, we are able to obtain a convergence rate \emph{without} modifying the subproblems.
\begin{theorem}\label{thm:CG-convergence}
	Let $C := \max_{v,w \in \cV} \sum_{s \in \cS} \frac{y_s}{n} (w(s) - v(s))^2$ which is finite since $\cV$ is compact. Let $v^* \in \cV$ be the optimal solution to \eqref{eq:non-parametric-finitedim} and $B \geq 1/2$ be chosen so that $2B - \sqrt{2B} = \gamma^2 (h(v^*) - h(\bar{v}_1))/C$. Let $\{\bar{v}_1,\ldots,\bar{v}_{\cT}\}$ be the iterates of \cref{alg:CG}. For $t \geq 1$, we have
	\[ h(v^*) - h(\bar{v}_{t+1}) \leq \frac{4 BC/\gamma^2}{t+2}. \]
\end{theorem}
\proof{Proof of \cref{thm:CG-convergence}.}
	Define $\bar{w}_t = \{ \bar{\lambda}_t^s \exp(-\bar{\lambda}_t) \}_{s \in \cS}$ and $w_\ell = \{ \lambda_{\ell}^s \exp(-\lambda_\ell) \}_{s \in \cS}$ for $\ell=1,\ldots,L$. Then each $\bar{v}_{t+1}$ can be viewed as maximizing $h(v)$ over $v \in \Conv\left( \{w_1,\ldots,w_L,\bar{w}_1,\ldots,\bar{w}_t\} \right)$. This shows that $h(\bar{v}_t) \leq h(\bar{v}_{t+1})$ for each iteration $t$.
	
	We will consider a proxy point $\tilde{v}_t = \bar{v}_t + \alpha (\bar{w}_t - \bar{v}_t)$, where $\alpha \geq 0$ is such that
	\[ \tilde{v}_t \geq \frac{\gamma}{\sqrt{2 B}} \bm{1}, \]
	and $B \geq 1/2$ is a constant to be chosen later.
	
	Applying a Taylor expansion of $h$ around $\bar{v}_t$ we get
	\begin{align*}
		h(\tilde{v}_t) &= h(\bar{v}_t) + \alpha \grad h(\bar{v}_t)^\top (\bar{w}_t - \bar{v}_t) + \frac{\alpha^2}{2} (\bar{w}_t - \bar{v}_t)^\top \grad^2 h(\bar{r}_t) (\bar{w}_t - \bar{v}_t)\\
		&= h(\bar{v}_t) + \alpha \grad h(\bar{v}_t)^\top (\bar{w}_t - \bar{v}_t) - \frac{\alpha^2}{2} \sum_{s \in \cS} \frac{y_s (\bar{w}_t(s) - \bar{v}_t(s))^2}{n r_t(s)^2}
	\end{align*}
	for some $\bar{r}_t = \bar{v}_t + \alpha_r (\bar{w}_t - \bar{v}_t)$, where $\alpha_r \leq \alpha$, i.e., $\bar{r}_t$ lies on the line segment between $\bar{v}_t$ and $\tilde{v}_t$. This means that $\bar{r}_t \geq \frac{\gamma}{\sqrt{2 B}} \bm{1}$ as well.
	
	Since $\bar{w}_t$ is chosen as the value $v$ which maximizes $\grad h(\bar{v}_t)^\top v$ we have $\grad h(\bar{v}_t)^\top (\bar{w}_t - \bar{v}_t) \geq \grad h(\bar{v}_t)^\top (v^* - \bar{v}_t)$. Thus concavity of $h$ implies that $\grad h(\bar{v}_t)^\top (\bar{w}_t - \bar{v}_t) \geq h(v^*) - h(\bar{v}_t)$ for the optimal $v^*$, and substituting in the bound for $\bar{r}_t$, we get
	\begin{align*}
		h(\tilde{v}_t) &\geq h(\bar{v}_t) + \alpha (h(v^*) - h(\bar{v}_t)) - \frac{\alpha^2 B}{\gamma^2} \sum_{s \in \cS} \frac{y_s (\bar{w}_t(s) - \bar{v}_t(s))^2}{n}\\
		&\geq h(\bar{v}_t) + \alpha (h(v^*) - h(\bar{v}_t)) - \frac{\alpha^2 B C}{\gamma^2}.
	\end{align*}
	Setting $\alpha = \frac{\gamma^2 (h(v^*) - h(\bar{v}_t))}{2 B C}$ maximizes the right hand side, but we need to ensure that $\tilde{v} _t$ satisfies the required bound. Thus
	\begin{align}
		\bar{v}_t + \frac{\gamma^2 (h(v^*) - h(\bar{v}_t))}{2 B C} (\bar{w}_t - \bar{v}_t) \geq \frac{\gamma}{\sqrt{2 B}} \bm{1}. \label{eq:CG-convergence-proof-bound}
	\end{align}
	Since $\bar{w}_t \geq 0$ and $\gamma \leq \bar{v}_t(s)$ for all $s \in \cS$, this holds if we can guarantee
	\[ 1 - \frac{1}{\sqrt{2 B}} \geq \frac{\gamma^2 (h(v^*) - h(\bar{v}_t))}{2 B C} \iff 2B - \sqrt{2B} \geq \frac{\gamma^2 (h(v^*) - h(\bar{v}_t))}{C}. \]
	Also, $h(\bar{v}_t) \geq h(\bar{v}_1)$, hence this holds if
	\[ 2B - \sqrt{2B} \geq \frac{\gamma^2 (h(v^*) - h(\bar{v}_1))}{C}. \]
	Since $2B - \sqrt{2B}$ is increasing for $B \geq 1/2$, there exists some $B \geq 1/2$ such that the above holds with equality, thus guaranteeing \eqref{eq:CG-convergence-proof-bound}.

	We also need $\alpha \le 1$. But our choice of $B$ guarantees that $1 > 1 - \frac{1}{\sqrt{2 B}} \geq \frac{\gamma^2 (h(v^*) - h(\bar{v}_t))}{2 B C}$, so that our choice of $\alpha$ satisfies this constraint. From the Taylor expansion above we have
	\begin{align*}
		h(\tilde{v}_t) &\geq h(\bar{v}_t) + \frac{\gamma^2 (h(v^*) - h(\bar{v}_t))^2}{4 B C}.
	\end{align*}
	Since $\alpha \in [0,1]$, we know that $\tilde{v}_t \in \Conv(\{\bar{w}_t, \bar{v}_t\}) \subset \Conv\left( \{w_1,\ldots,w_L,\bar{w}_1,\ldots,\bar{w}_t\} \right)$, thus by definition of $\bar{v}_{t+1}$ we have $h(\tilde{v}_t) \leq h(\bar{v}_{t+1})$, hence
	\[ h(\bar{v}_{t+1}) \geq h(\bar{v}_t) + \frac{\gamma^2 (h(v^*) - h(\bar{v}_t))^2}{4 B C} \]
	Adding $h(v^*)$ to both sides, defining $\Delta_t := h(v^*) - h(\bar{v}_t)$, and rearranging, we have the recurrence relation
	\[ \Delta_{t+1} \leq \left( 1-\frac{\gamma^2 \Delta_t}{4 B C} \right) \Delta_t. \]
	This can be rewritten
	\[ \frac{1}{\Delta_t} + \frac{\gamma^2}{4BC - \gamma^2 \Delta_t} \leq \frac{1}{\Delta_{t+1}}. \]
	Thus
	\[ \frac{1}{\Delta_t} + \frac{\gamma^2}{4BC} \leq \frac{1}{\Delta_{t+1}}. \]
	This implies $\frac{1}{\Delta_{t+1}} \geq \frac{1}{\Delta_{1}} + \frac{\gamma^2t}{4BC}$, and hence
	\[ \Delta_{t+1} \leq \frac{4 \Delta_1 BC}{4BC + \gamma^2 \Delta_{1} t  } = \frac{4BC/\gamma^2}{4BC/(\gamma^2 \Delta_1) + t}. \]
	By our choice of $B$ we have $1 - \frac{1}{\sqrt{2B}} \geq \frac{\gamma^2 \Delta_1}{2BC}$ which means $\frac{\gamma^2 \Delta_1}{4BC} < \frac{1}{2}$ and $4BC/(\gamma^2 \Delta_1) > 2$. Thus we have
	\[ \Delta_{t+1} \leq \frac{4BC/\gamma^2}{t+2} \]
	as required.
\qed
\endproof

\section{Further numerical results}\label{app:conjugate-prior}

We ran some simple tests using a gamma prior $\lambda \sim g=\GammaDist(\alpha,\theta)$ and $X,\xi \mid \lambda \sim \Poisson(\lambda)$. Since the gamma distribution is conjugate to the Poisson distribution, thus the marginal and posterior distributions have closed forms, which are negative binomial: $X \sim \NB\left( \alpha, \frac{1}{1+\theta} \right)$ and $\xi \mid X \sim \NB\left( \alpha + X, \frac{1+\theta}{1+2\theta} \right)$. We can thus compare the exact posterior probability $\bbP[\xi=k \mid X]$ with the one computed using the exact marginal distribution $f$ of $X$ given by the generalized Robbins formulae \eqref{eq:post_prob_f}.

However, comparing estimates of posterior probabilities obtained via \eqref{eq:post_prob_f} with the actual posterior probabilities $\bbP[\xi=k \mid X]$ shows that numerical error is introduced, even when the exact marginal is available. We tabulate this for $X = 8$ in \cref{tab:full-plugin-instability}, and see that estimates for $k\geq 10$ start to degrade badly, and can even be negative.


\begin{table}[htp]
	\centering
	\caption{Comparing generalized Robbins formulae versus exact posterior probabilities when $\lambda \sim \text{Gamma}(\alpha,\theta)$ with $\alpha=2$, $\theta=2$.}\label{tab:full-plugin-instability}
	\begin{tabular}{r|r|r}
		\toprule
		$k$ &   $\bbP[\xi=k \mid X=8]$ &  est. $\bbP[\xi=k \mid X=8]$ via \eqref{eq:post_prob_f} \\
		\midrule
		0 & 0.006047 &   0.006047 \\
		1 & 0.024186 &   0.024186 \\
		2 & 0.053210 &   0.053210 \\
		3 & 0.085136 &   0.085136 \\
		4 & 0.110677 &   0.110677 \\
		5 & 0.123959 &   0.123960 \\
		6 & 0.123959 &   0.123951 \\
		7 & 0.113334 &   0.113285 \\
		8 & 0.096334 &   0.096069 \\
		9 & 0.077067 &   0.076412 \\
		10 & 0.058571 &   0.057512 \\
		11 & 0.042597 &   0.046180 \\
		12 & 0.029818 &  -0.012966 \\
		13 & 0.020184 &  -0.108988 \\
		14 & 0.013264 &  -0.432127 \\
		15 & 0.008489 &   0.453766 \\
		\bottomrule
	\end{tabular}
\end{table} 

\end{appendices}

\bibliographystyle{abbrvnat}
\bibliography{bib,new}

\end{document}